\documentclass{cta-author}

\usepackage{color}
\usepackage{xcolor}
\usepackage{lineno}
\usepackage{hyperref}
\usepackage{booktabs}
\usepackage{graphicx}
\usepackage{subfigure}
\usepackage{epstopdf}
\usepackage{picinpar}
\usepackage{amsmath}
\usepackage{amsfonts}
\usepackage[mathscr]{euscript}
\usepackage{calligra}
\usepackage{array}

{}
{}
\newtheorem{remark}{Remark}{}

\begin{document}

\supertitle{Submission Template for IET Research Journal Papers}

\title{Coordinated Formation Control for Intelligent and Connected Vehicles in Multiple Traffic Scenarios}

\author{\au{Qing Xu$^{1}$}, \au{Mengchi Cai$^{1}$}, \au{Keqiang Li*$^{1}$}, \au{Biao Xu$^{1}$},\\ \au{Jianqiang Wang$^{1}$} and \au{Xiangbin Wu$^{2}$}}

\address{\add{1}{School of Vehicle and Mobility, Tsinghua University, Beijing, China}
\add{2}{Intel Lab China}
\email{likq@tsinghua.edu.cn}}

\begin{abstract}
In this paper, a unified multi-vehicle formation control framework for Intelligent and Connected Vehicles (ICVs) that can apply to multiple traffic scenarios is proposed. In the one-dimensional scenario, different formation geometries are analyzed and the interlaced structure is mathematically modelized to improve driving safety while making full use of the lane capacity. The assignment problem for vehicles and target positions is solved using Hungarian Algorithm to improve the flexibility of the method in multiple scenarios. In the two-dimensional scenario, an improved virtual platoon method is proposed to transfer the complex two-dimensional passing problem to the one-dimensional formation control problem based on the idea of rotation projection. Besides, the vehicle regrouping method is proposed to connect the two scenarios. Simulation results prove that the proposed multi-vehicle formation control framework can apply to multiple typical scenarios and have better performance than existing methods.
\end{abstract}

\maketitle

%
\section{Introduction}
\label{section1}
%

With the development of V2X communication, coordinated control of ICVs is attracting great attention. Compared with conventional vehicle control, coordinated control of multiple vehicles focuses on the payoff of group of vehicles instead of single vehicle.

There have been numerous research on the topic of vehicular coordinated control. Traffic scenarios for vehicular coordinated control can be categorized into one-dimensional and two-dimensional methods. In this paper, the one-dimensional scenario is defined as multi-lane scenarios where vehicles drive with the same direction, where the direction mainly indicates macroscopic direction in the road coordinate, although individual vehicles may have different orientation when they are changing lanes or driving on a curve. Typical one-dimensional scenarios include straight or curve multi-lane road segment. Intersections and on/off-ramps are typical two-dimensional scenarios.

Existing research regards the coordinated control of ICVs at one-dimensional scenarios as typical examples of convoy control or formation control. On one hand, the existing research of multi-vehicle formation control can be macroscopically divided into structuralized-road scenarios,
 \cite{2qian2016hierarchical, 3kato2001cooperative, 4navarro2016distributed, 5marjovi2015distributed, 6kato2002vehicle, 7marinescu2010active, hall1999design, 21Dao2007Optimized, 8marinescu2012ramp,li2020optimal,2019Multi},
 and unstructuralized-road scenarios,
 \cite{9liu2014dynamic,10gowal2010local}.
For the structuralized-road scenario, a group of vehicles within a given range will form a formation and drive coordinately. The geometric shape of the formation is built under a certain rule, and the shape changes when the number of the lanes changes,
 \cite{3kato2001cooperative, 6kato2002vehicle, 7marinescu2010active},
 or when there are some obstacles on the road,
 \cite{2qian2016hierarchical, 5marjovi2015distributed}.
Formation adjusting functions when vehicles join or leave the convoy are discussed in
 \cite{5marjovi2015distributed}
 and
 \cite{8marinescu2012ramp}.
 And the coordinated lane change behavior in a built formation is studied in
 \cite{4navarro2016distributed}.
The methods to cooperatively control ICVs together with human-driving vehicles are studied in 
 \cite{li2020optimal}
 and
 \cite{2019Multi}.
Other research about the coordinated behavior of multiple ICVs can be seen in 
 \cite{ hall1999design, 21Dao2007Optimized, 8marinescu2012ramp,2020Ethical}.
Besides the research above which considers the multi-vehicle formation problem from the overall perspective, there is also some research which considers this problem from single-vehicle perspective,
 \cite{11kamal2016efficient}.
On the other hand, compared with vehicular formation control which has complex kinematic and dynamic constraints, there are much more research on robotic formation control which has simpler constraints. The robotic formation control, also known as swarm control, can be generally studied in the following three application areas: the autonomous grounded vehicles (AGV),
 \cite{12rezaee2014decentralized, 13han2017multi, 14macdonald2011multi},
 the unmanned aerial vehicles (UAV),
 \cite{15alonso2015collision, 16chen2015decoupled},
 and the autonomous underwater vehicles (AUV),
 \cite{17millan2014formation, 18park2015adaptive}.

The method proposed in this paper is inspired by the robotic control methods.

The coordinated control methods of ICVs in two-dimensional scenarios concentrate on intersections and ramps. As discussed in 
 \cite{Xu2018Distributed},
the traffic signal at intersections may bring the green time loss, which results in inefficiency for intersection management. Therefore, coordinated control methods of ICVs at unsignalized intersections is a potential solution for improving traffic efficiency. As a classical method which solves vehicle merging problem at on-ramps, the virtual platoon method is also an adoptable method for unsignalized intersections,
 \cite{li2006cooperative, Xu2018Distributed}.
By projecting vehicles on different legs of the intersection to a virtual platoon, the two-dimensional vehicle merging problem is transformed into one-dimensional platoon control problem. Besides of the virtual platoon method, another well-known method for the unsignalized intersections is the reservation method,
 \cite{dresner2008multiagent, huang2012assessing, zhang2015state},
 where vehicles adjust their speed and pass the intersections without collision according to the ``First Come, First Served" policy. Other methods developed for the unsignalized intersections can be found in 
 \cite{lee2012development},
\cite{mirheli2019consensus}
 and 
 \cite{li2019temporal} .

Here the shortcomings of the existing methods are summarized:
\begin{itemize}
\item The existing methods for different scenarios are often separated and lack connectivity and generality. There are only a few recent works that have taken the problem of multi-scenarios into account or have proposed unified decision and control methods.

\item The scenarios are often over simplified for vehicular coordinated control. For example, the classic virtual platoon method can only be applied to single-lane ramp merging scenario.

\item Some methods have only focused on single vehicle and analyzed its payoff instead of vehicle formations' or sub-groups', which could not fully utilize the advantage of vehicular coordinated control.

\item  The desired assignment of the vehicles in a formation is often set beforehand and lacks principle explanation, which makes it hardly to be applied to the real world. Besides, the choosing of formation geometry is kind of arbitrary and lacks reasonable analysis among the existing research.
 
\item Although there are some preliminary researches of robotic formation control, they can be hardly directly applied to vehicular formation control because of the lack of kinematic constraints.

\end{itemize}

This paper studies the problem of coordinated control of ICVs in multiple traffic scenarios. In the one-dimensional scenario, a multi-lane formation control method is proposed to form vehicular formations. In the two-dimensional scenario, an improved virtual platoon method is proposed to calculate conflict-free passing sequence of vehicles at multi-lane unsignalized intersections. The vehicle regrouping method is proposed to divide the vehicular formations into sub-groups that can pass the intersection ordered and with high efficiency.

The rest of this paper is organized as follows. Section \ref{section2} sets up the scenarios studied in this paper and provide vehicle controllers. Section \ref{section3} presents the formation control method for the one-dimensional scenario. Section \ref{section4} presents the improved virtual platoon method for the two-dimensional scenario. The vehicle regrouping method is proposed to connect the two scenarios in Section \ref{section5}. Simulation experiments and results are given in Section \ref{section6}. At last, this paper is concluded in Section \ref{section7}.

%
\section{System modeling}
\label{section2}
%

This paper considers a traffic system with multiple ICVs and multiple traffic scenarios. All the vehicles in this system are ICVs, and are both controllable and observable. The state feed-back of all the vehicles is supposed to be available and the quality of communication is supposed to be perfect. The model, parameters and controller of typical traffic scenarios and vehicles studied in this paper are discussed in this section.

\subsection{Traffic scenarios}

\begin{figure}
\vspace{3mm}
\begin{center}
    \subfigure[One-dimensional highway segment]{
    \includegraphics[scale = 0.27]{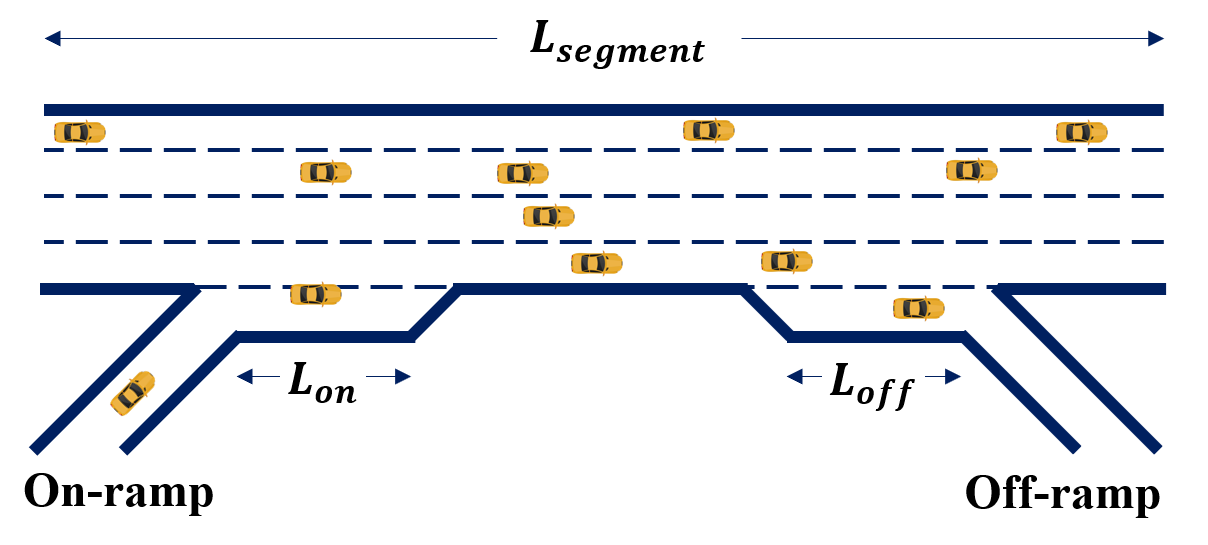}
    \label{fig_envir}}
\hspace{5mm}
    \subfigure[Two-dimensional typical intersection]{
    \includegraphics[scale = 0.44]{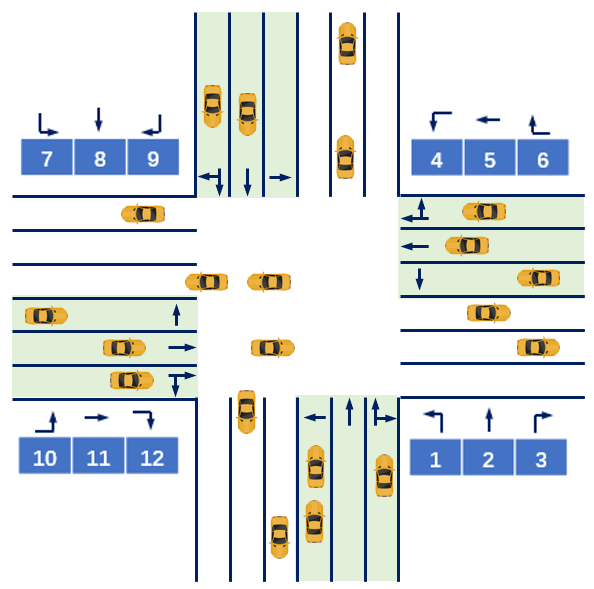}
    \label{tflow}}
    \caption{Scenarios studied in this paper.}
    \label{figure1}

\end{center}
\end{figure}

As defined in Section \ref{section1}, this paper studies coordinated control methods for one-dimensional and two-dimensional traffic scenarios. The methods to be proposed in this paper are designed for multiple scenarios, but in order to detailedly explain our methods and algorithms, the multi-lane highway system with on-ramps and off-ramps and the multi-lane unsignalized intersection are chosen to be the basic scenarios of this paper. An isolated highway segment with an on-ramp and an off-ramp is shown in Fig. \ref{fig_envir}. Vehicles come from the on-ramp or the last segment and drive on the main road. The on-ramp and off-ramp are located on the rightside of the main road. More complex highway systems can be considered as combination of several simple road segments. The unsignalized intersection studied in this paper is shown in Fig. \ref{tflow}. In this paper, for simplicity, a symmetric multi-lane intersection which have 6 lanes of each leg, 3 approaching-lanes and 3 departure-lanes, is considered. Each approaching-lane consists of one turning-left lane on the left, one going-straight and turning-right lane in the middle and one turning-right lane on the right. Vehicles from different directions of the intersection can turn left, go straight and turn right, forming 12 potential traffic flow movements at this intersection. The traffic flow movements are numbered with 1–12.

This paper aims to provide a coordinated control method which is not sensitive with scenarios. However, it's obvious that when the lane assignment of the intersection changes, the specific implementation process of the method also changes. Thus, the typical intersection scenario provided in Fig. \ref{tflow} is chosen for this paper to clearly explain the proposed method.

\subsection{Vehicle model, parameters and controller}

\begin{figure}
\begin{center}
    \includegraphics[scale = 0.5]{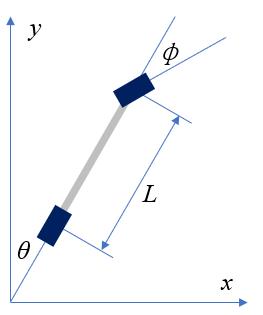}
    \caption{Bicycle model. $L$ represents the wheelbase, $\phi$ represents the steering angle of the front wheel and $\theta$ represents the yaw angle. }
    \label{fig_bicyclemodel}
\end{center}
\end{figure}

In order to simplify computation, the bicycle model is used in this paper, as is shown in Fig. \ref{fig_bicyclemodel}. The vehicle parameters used in this paper are listed in Table \ref{tab_vehiclepara} where $v$ represents the longitudinal speed, $a$ represents the acceleration and $\phi$ represents the steering angle.

\begin{table}
\centering
\caption {Vehicle parameters used in this paper.}
\label {tab_vehiclepara}
\begin{tabular}{|c|c|c|c|c|c|}
\hline
$L$ & $\phi$ & $\theta$ & $v$ & $a$ \\
(m) & (degree) & (degree) & (m/s) & (m/$\text{s}^2$) \\
\hline
$2.8$ & $[-30,30]$ & $[-180,180)$ & $[10,30]$ & $[-6,3]$ \\
\hline
\end{tabular}
\end{table}

The state of the vehicle contains $X$, $Y$, $\theta$ and $\phi$, where $X$ and $Y$ represent the horizontal and longitudinal position of the center of vehicle's rear axle. The inputs of the vehicle model are $a$ and $\phi$, and the state is calculated as:
\begin{eqnarray}
\begin{cases}
\label{eq_state}
\dot{X}=v\cdot sin(\theta)\\
\dot{Y}=v\cdot cos(\theta)\\
\dot{\theta}=tan(\phi)\cdot v/L\\
\dot{v}=a\\
\end{cases}
\end{eqnarray}

In this paper, the control strategy of vehicles fuses the control inputs of both formation maintaining and desired speed keeping. The longitudinal controller is designed to form and maintain the generated formation and keep a desired speed simultaneously. When the vehicle is driving on the straight road, the forward direction is chosen as the positive direction of X-axis and the positive direction of Y-axis pointe to the left of the vehicle. The longitudinal feedback control law of vehicles is given as:
\begin{eqnarray}
\label{eq_law}
a_i=k_{pi}(X_i-X_{\text{ref,}i})+k_{vi}(v_i-v_{\text{ref,}i})
\end{eqnarray}
where $a_i$ is the control input of vehicle $i$ and $k_{pi}$ and $k_{vi}$ are feedback gains. $X_{\text{ref,}i}$ and $v_{\text{ref,}i}$ are reference position and speed of the vehicle. 

The lateral controller is designed to perform lane-keeping and lane-changing manoeuvre. 3-order B$\acute{\text{e}}$zier curve is used for trajectory generating, and the preview method is chosen to guide vehicle to follow the trajectory.

%
\section{Formation control in one-dimensional scenarios}
\label{section3}
%

This section introduces the  multi-lane formation control method under the one-dimensional scenario. As is previously discussed in \cite{cai2019multi}, the one-dimensional formation control problem is solved in the following three steps: generating of target positions, assigning vehicles to target positions and target-following control of vehicles.

\subsection{Geometric structure of multi-lane formation}

First of all, the standard formation geometry of vehicle formations on the multi-lane straight road is defined.

Two typical types of formation geometries, the interlaced formation and the parallel formation, are commonly chosen in the existing research. As is shown in Fig. \ref{figure1}, although the interlaced formation reduces the vehicle density, it provides great convenience for lane changing and formation adjusting. Although the parallel formation has higher vehicle density, the flexibility of formation adjusting and the safety of coordinated driving are not as good as the interlaced formation. This paper mainly focuses on the real-time formation generating and adjusting, and the adjusting flexibility of formation is highly required when the number of lanes or number of vehicles changes. Thus, the interlaced formation geometry is adopted for multi-lane formation. The details of the interlaced formation can be found in 
 \cite{2qian2016hierarchical},
 \cite{3kato2001cooperative},
 \cite{5marjovi2015distributed},
 \cite{6kato2002vehicle}
 and
 \cite{8marinescu2012ramp}.
The parallel formation is chosen in 
 \cite{7marinescu2010active}.

\begin{figure}
\vspace{3mm}
\begin{center}
    \subfigure[Interlaced geometry]{
    \includegraphics[scale = 0.55]{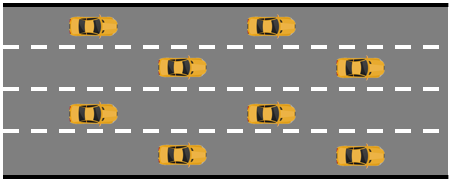}
    \label{formation1}}
\hspace{5mm}
    \subfigure[Parallel geometry]{
    \includegraphics[scale = 0.55]{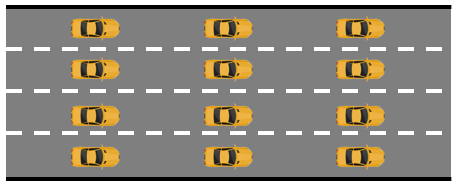}
    \label{formation2}}
    \caption{Different types of formation. (a) shows the interlaced formation where vehicles in adjacent lanes drive interlacedly in longitudinal direction; (b) shows the parallel formation where vehicles drive parallelly in longitudinal direction.}
    \label{figure1}
\end{center}
\end{figure}

\begin{figure}
\begin{center}
    \subfigure[4 vehicles]{
    \includegraphics[scale = 0.2]{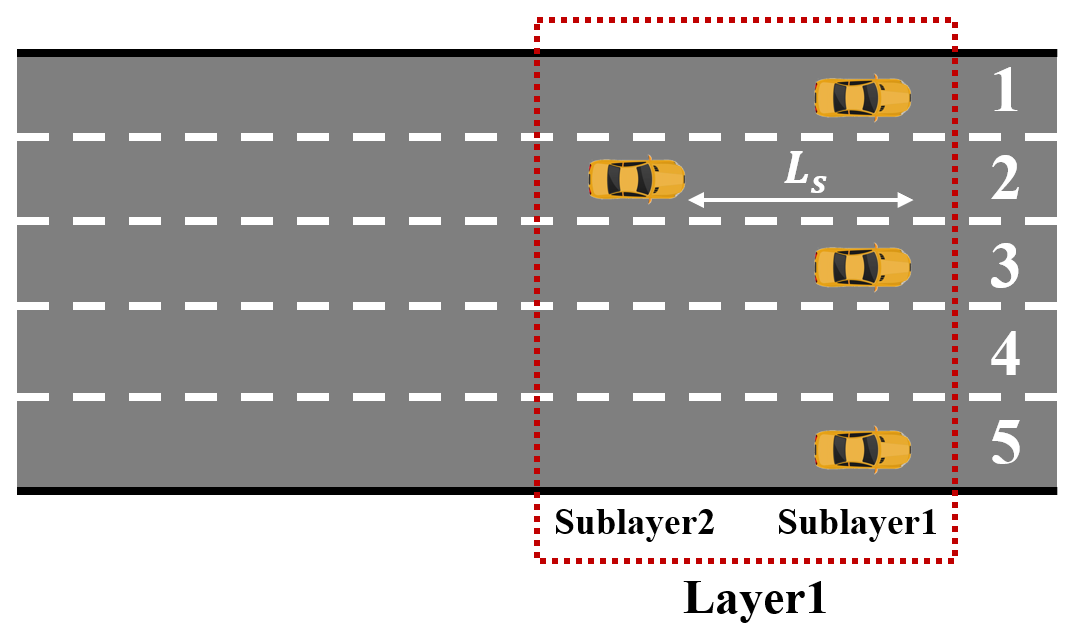}
    \label{figure3_1}}
    \subfigure[7 vehicles]{
    \includegraphics[scale = 0.2]{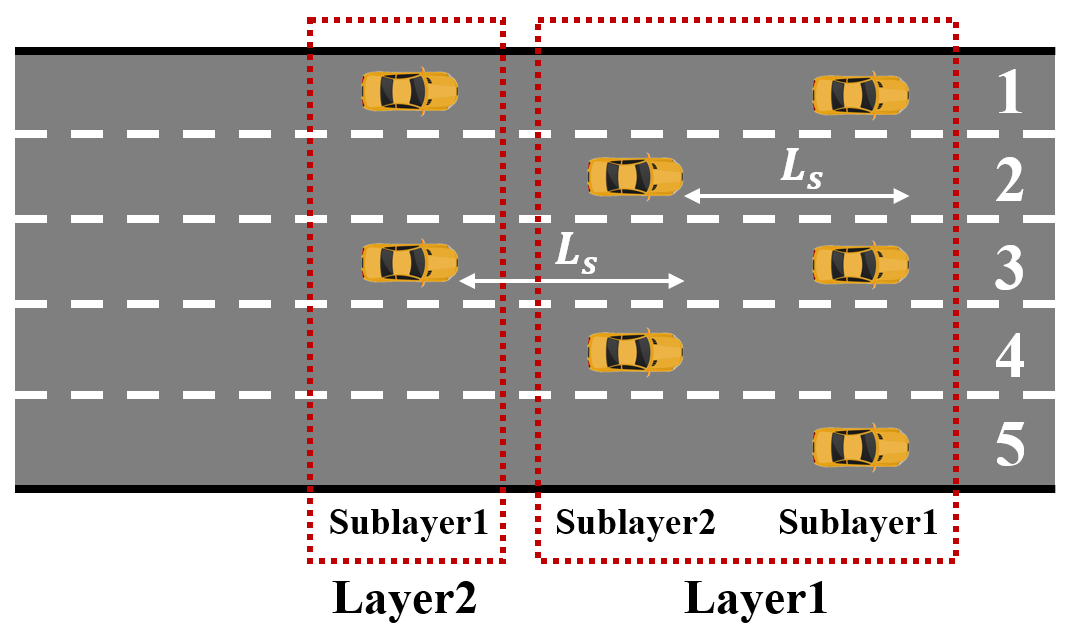}
    \label{figure3_2}}
    \subfigure[9 vehicles]{
    \includegraphics[scale = 0.2]{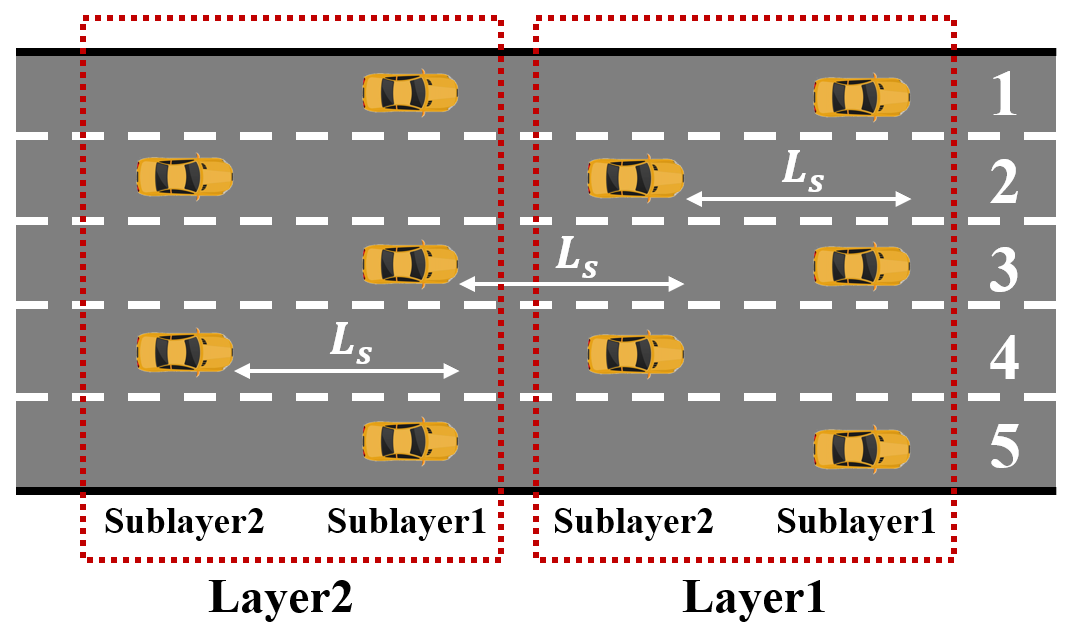}
    \label{figure3_3}}
    \caption{Layered structure of the multi-lane formation. Each layer is then divided into two sublayers and vehicles in adjacent lanes are in different sublayers. In (a) there is only one layer that contains two sublayers; In (b) and (c) there are two layers and the first layers are both filled up. The second layer in (b) contains one sublayer and the second layer in (c) contains two sublayers.}
    \label{figure3}
\end{center}
\end{figure}

In order to make the formation repeatable and the positions of targets inside the formation descriptive, targets are divided into different layers
based on their longitudinal position. In each layer, there is at most one target in each lane, so the maximum number of targets equals to the
number of lanes. The targets in the same layer are divided into two sublayers and targets in the same sublayer keep the same longitudinal position
and the lanes they occupy are not adjacent. The two sublayers in the same layer keep a longitudinal distance of {\it d} and the two adjacent layer
also keep a longitudinal distance of {\it d}. The layered structure of multi-lane formation is shown in Fig. \ref{figure3}.

\subsection{Target generating rules for targets in multi-lane formation}

This subsection provides the formation generating rule to generate targets in the multi-lane formation in real time. Suppose there are $N_{l}$ lanes, and their ID are numbered 1 to $N_{l}$ from left to right. The coordinate $(x_g,y_g)$ is used to define the relative position of a target in a formation where $x_g$ represents the reference longitudinal distance between this target and the most forward target and $y_g$ represents the lane ID where the target locates. The $i$-th target position in the formation can be calculated based on the following rules:
\begin{eqnarray}
\begin{cases}
\label{xy}
x_{gi}=&(P_2(\text{ceil}(\frac{i}{N_{l}}))\\&+\text{com}(\text{mod}(i,N_{l}),\text{ceil}(\frac{N_{l}}{2})))\times L_s\\
y_{gi}=&P_1(\text{}mod(i,N_{l}))-\text{com}(P_1(\text{mod}(i,N_{l})),\\&\text{ceil}(\frac{N_{l}}{2}))\times P_1(\text{ceil}(\frac{N_{l}}{2}))
\end{cases}
\end{eqnarray}

where $L_s$ represents the longitudinal distance between two adjacent sublayers. The distance between two sublayers in one layer is defined the same as the distance between two adjacent layers. The two polynomial functions $P_1(p)$ and $P_2(p)$ are defined as:

\begin{eqnarray}
\label{eq1}
P_1(p)=2\times p-1\\
P_2(p)=2\times p-2
\end{eqnarray}

The function mod$(p,q)$ returns the remainder of $q$ divided by $p$. The function ceil$(p)$ returns the minimum integer no less than $p$. The comparing function com$(p,q)$ is defined as:
\begin{eqnarray}
\text{com}(p,q)=
\begin{cases}
1 & \text {if  } p > q \\
0 & \text {otherwise}
\end{cases}
\end{eqnarray}

Take the first target position locating at the most left-forward of the formation as an example. Its coordinate can be calculated by (\ref{xy}) and it is $(1,0)$. The number of target positions equals to the number of vehicles. When the road has 5 lanes and there are different numbers of vehicles on it, the target positions can be calculated and shown in Fig. \ref{figure3}.

\subsection{Position assignment}

Vehicles are assigned to the generated target positions in this step. There are $N!$ different assignment results when the number of vehicles is $N$. By defining function $F(i,j)$ to represent the cost for vehicle $i$ to be assigned to target $j$, the total cost of every assignment result can be calculated, and $F(i,j)$ is called the cost function. Accordingly, the cost matrix is defined as:
\begin{eqnarray}
 \mathcal{C}=[c_{ij}]\in \mathbb{R}^{N\times N},\ \ c_{ij}=F(i,j),\ \ i,j\in \mathbb{N}^+
\end{eqnarray}

The assignment matrix $\mathcal{A}$ is defined to represent the result of any specific assignment:
\begin{center}
\begin{eqnarray}
&\mathcal{A}=[a_{ij}]\in \mathbb{R}^{N\times N} \\
&a_{ij}=
\begin{cases}
1 \ \text{if vehicle $i$ is assigned to target $j$}\notag \\
0 \ \text{otherwise}
\end{cases}
\\ &i,j\in \mathbb{N}^+\notag
\end{eqnarray}
\end{center}

And the assignment problem can be modeled as a 0-1 integer programming problem:
\begin{alignat}{2}
\min\quad & \sum_{i=1}^N\sum_{j=1}^N (c_{ij}\times a_{ij}) &{}& \label{eqn - lp}\\
\mbox{s.t.}\quad
&\sum_{i=1}^N a_{ij}=1\notag \\
&\sum_{j=1}^N a_{ij}=1\notag \\
&i,j\in \mathbb{N}^+\notag
\end{alignat}
where $[c_{ij}]$ is a given cost matrix and $[a_{ij}]$ is the variable assignment matrix.

In this research, in order to consider lateral lane change safety and longitudinal following performance, the cost function $F_w(i,j)$ is constructed as a weighted summation of longitudinal distance and lateral lane change numbers. The lane change number is chosen as part of the cost function to represent lateral movement of the vehicle because the target positions are always located on the center of the lanes. The cost function $F_w(i,j)$ has the following structure:
\begin{eqnarray}
F_w(i,j)=w_1(x_{gj}-x_{vi})^2+ w_2(y_{gj}-y_{vi})^2
\end{eqnarray}
where $x_v$ and $y_v$ represent the longitudinal position and lane ID of a vehicle. By choosing different weight $w_1$ and $w_2$, the cost function $F(i,j)$ changes its preference of longitudinal distance or lateral distance between vehicles and target positions. For example, when $w_2$ is set as the width of the lane and $w_1$ is set 1, then $F_w(i,j)$ represents the square of Euclidean distance. Complex parameter tuning methods are not adopted because we focus on the assignment framework designing, and different tunning methods can be chosen for different application scenarios.

\begin{figure}
\begin{center}
    \includegraphics[scale = 0.24]{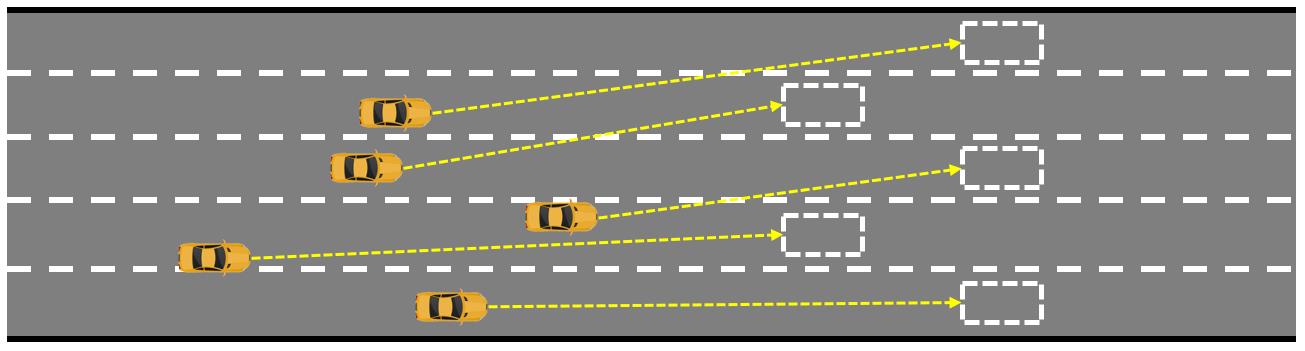}
    \caption{Example result of Hungarian algorithm using Euclidean distance as the assigning cost between vehicles and targets. The white rectangles represent the targets and the yellow dashed lines represent the assignment relationship between vehicles and targets.}
    \label{figure4}
\end{center}
\end{figure}

Among all the possible assignment results, the one with the minimum total cost is chosen as the best assignment. The Hungarian algorithm,
 \cite{19kuhn1955hungarian,20munkres1957algorithms},
is chosen to solve the assignment problem for its low time complexity ($O(N^3)$). Other research which uses the Hungarian method for assignment problem includes \cite{14macdonald2011multi}. An example result of using Euclidean distance as the cost is shown in Fig. \ref{figure4}.

\subsection{Vehicle control}

After vehicles are assigned to target positions, trajectories are generated to guide vehicles towards their expected target positions. In this paper, the 3-order B$\acute{\text{e}}$zier curve is chosen to be the candidate trajectory of lane change. The coordinate of the target position is used as inputs of the vehicle afterwards. For more detailed information about using B$\acute{\text{e}}$zier curves to perform trajectory planning, please turn to \cite{gonzalez2014continuous}. 

Trajectory following and collision avoidance are also studied in this research, but this paper focuses on the upper-level decision making and control. For more information about the lower-level controller and hardware architecture, please turn to our previous conference paper \cite{chen2019system}.

In this paper, the obstacle avoidance problem is considered as: if the obstacle does not occupy enough space to block a lane for vehicles to drive, the formation doesn't change and the vehicle just need adjust its trajectory within the lane; if the obstacle causes the lane it occupied undrivable, this case will be the same as the case in which the number of lanes changes, and can be treat using the same method. Besides, it is easy to extend the straight-road method to the curve roads and this paper focuses on the adjusting functions of the formation in the straight-road scenario.

%
\section{Virtual platoon method in two-dimensional scenarios}
\label{section4}
%

The two-dimensional scenarios include on-ramps, off-ramps, roundabouts and intersections. The two-dimensional vehicular formation control method proposed in this paper is scenario-insensitive and can be applied to all the mentioned scenarios. In order to make it easier for readers to understand the method, this paper focuses on the unsignalized multi-lane intersections. This section introduces the virtual platoon method used for unsignalized single-lane intersections and then apply it to the multi-lane scenarios.

\subsection{Traditional virtual platoon method}

The virtual platooning method is a well-known method to handle the two-dimensional traffic scenarios. Typical applications are on-ramp merging control and unsignalized intersection control. The main idea of virtual platooning method is to project vehicles on different lanes to a virtual lane according to the distance between vehicles and the center of the intersection. Then the single-lane platooning control method can be applied to the vehicles on the virtual lane. Thus, vehicles form a platoon without longitudinal conflict so that they can pass the intersection smoothly without conflict. Other research using virtual platoon method can be seen in 
 \cite{li2006cooperative}.

\subsection{Improved virtual platoon method}

In the multi-lane intersections, vehicles are treat with a similar idea of the single-lane intersections. In the single-lane scenario, each single vehicle is treated as a virtual vehicle in the virtual platoon. However, in the multi-lane scenario, a group of vehicles with the same target direction (turning left, going straight or turning right) are treated as a node in the virtual platoon. If these nodes are put in a single-lane platoon, they can pass the intersection without braking. In order to improve traffic efficiency, different nodes (vehicle group) whose target directions are not in conflict can pass the intersection simultaneously. We first proposed this idea in \cite{Xu2018Distributed}. In order to get a tighter virtual platoon, conflict model of the traffic flow movements is built. 

\begin{figure}
\begin{center}
    \subfigure[Converging]{
    \includegraphics[scale = 0.25]{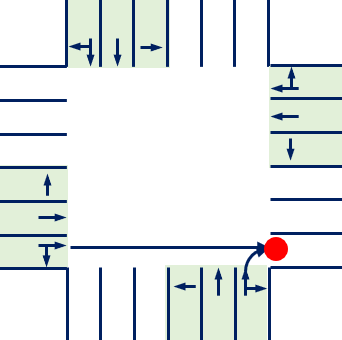}
    \label{converge}}
\hspace{1mm}
    \subfigure[Diverging]{
    \includegraphics[scale = 0.25]{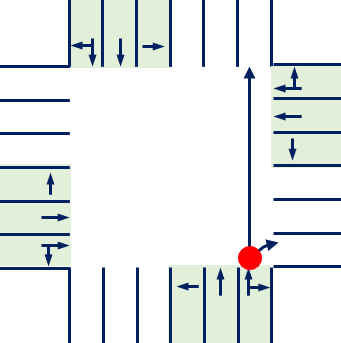}
    \label{diverge}}
\hspace{1mm}
    \subfigure[Crossing]{
    \includegraphics[scale = 0.25]{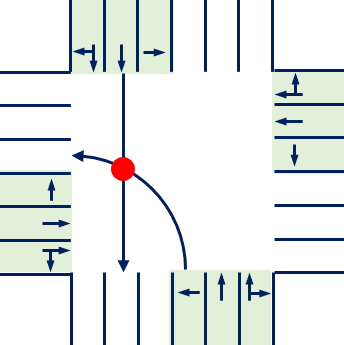}
    \label{cross}}
    \caption{Basic conflict model of traffic flow movements.}
    \label{figure_conflict}
\end{center}
\end{figure}

There are three types of conflict in an intersection: converging, diverging and crossing, as shown in Fig. \ref{figure_conflict}. The converging conflict happens when two kinds of traffic flow merge into the same lane. The diverging conflict happens when two kinds of traffic flow come from the same lane. It is important to notice that two vehicles coming from the same direction are always in conflict in the single-lane intersections, but not always in the multi-lane intersections because they may come from different lanes. The crossing conflict happens when the trajectories of two kinds of traffic flow cross in the intersection. The two turning-left traffic flow from the opposite direction are defined not in conflict, which can be achieved by special designing of intersection structure.

Compared with setting one vehicle group as a node in the virtual platoon, the traffic efficiency can be further improved by putting some nodes in the same tree-layer of the virtual platoon. The conflict relationship of traffic flow of a defined intersection is closely related to its lane assignment. The unsignalized multi-lane intersection studied in this paper is already shown in Fig. \ref{tflow}.

\begin{remark}
In order to distinguish from the definition given in Section.\ref{section3}, the layers of the depth-first spanning tree are named as tree-layers.
\end{remark}

According to our previous work, two zones are necessarily defined: approaching zone and coordinating zone. Additionally, the adjusting zone is specially defined for multi-lane intersections. In the approaching zone, vehicles follow the multi-lane formation control method to keep a stable group. In the adjusting zone, vehicles perform coordinated lane change to drive on their expecting lanes and re-form smaller formations. In the coordinating zone, the formations in which vehicles have the same destination are taken as a node and use the virtual platoon control method to guide them to pass the intersection. Sets $\mathbb{S}_{app}$, $\mathbb{S}_{adj}$ and $\mathbb{S}_{cor}$ are defined to represent sets containing vehicles in the approaching zone, adjusting zone and coordinating zone respectively. $R_1$ and $R_2$ are defined to be the radius of the boundary circle of the coordinating zone and the adjusting zone. Given a vehicle $veh_i$ which is approaching the intersection and its distance to the center of the intersection $d_i$, the state of the vehicle is decided according to the following rules:
\begin{eqnarray}
\begin{cases}
\label{eq6}
veh_i\in \mathbb{S}_{app},\   d_0>R_2\\
veh_i\in \mathbb{S}_{adj},\   R_2\geq d_0>R_1\\
veh_i\in \mathbb{S}_{cor},\   R_0\leq d_1\\
\end{cases}
\end{eqnarray}

\begin{figure}
\vspace{2mm}
\begin{center}
    \subfigure[Conflict relationships]{
    \includegraphics[scale = 0.35]{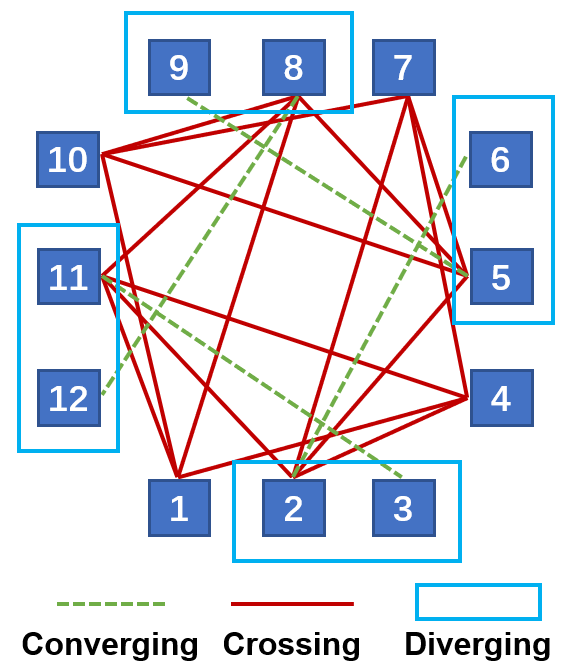}
    \label{graph2}}
    \subfigure[Direct graph]{
    \includegraphics[scale = 0.175]{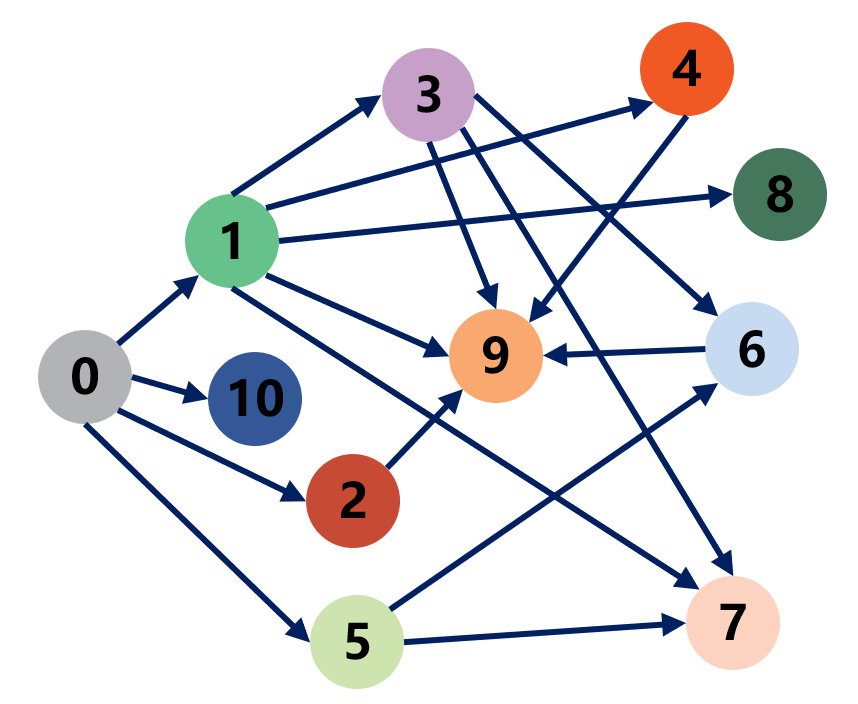}
    \label{direct2}}
    \subfigure[Depth-first spanning tree]{
    \includegraphics[scale = 0.175]{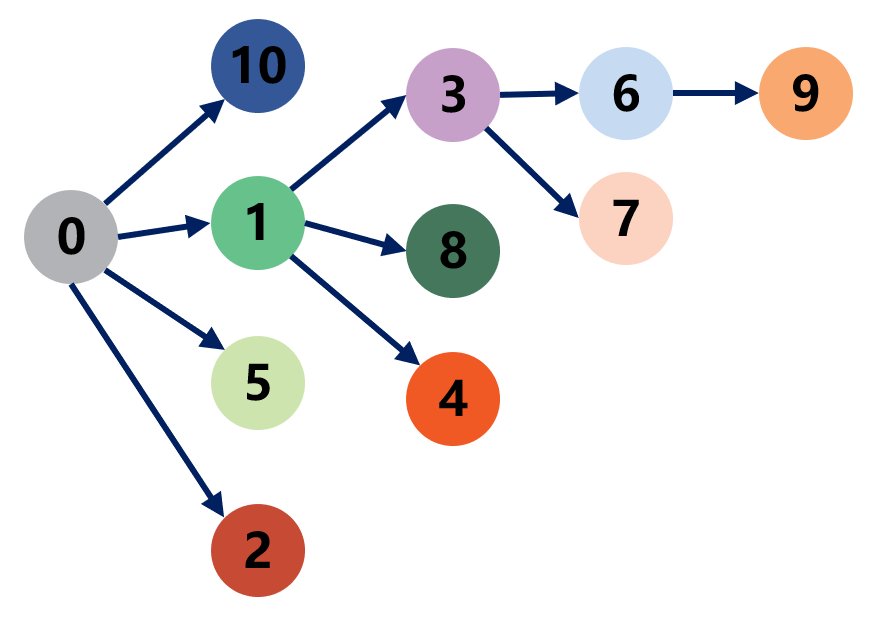}
    \label{tree2}}
    \caption{Process of calculating passing sequence for unsignalized multi-lane intersections.}
    \label{figure_single_inter}
\end{center}
\end{figure}

Because of the specific structure of the multi-lane intersection, the conflict relationship is different with the single-lane intersections. For example, the turning-left traffic flow and the going-straight traffic flow are in conflict in the single-lane intersection, but in the multi-lane intersection they are not in conflict because they use different lanes. Thus, a conflict relationship model is built for this specific multi-lane intersection, as shown in Fig. \ref{graph2} ,where the numbers in the blue rectangles represent the traffic flow movements which have been defined in Fig. \ref{tflow}, dashed lines represent the converging relationship, solid lines refer to the cross relationship, and the hollow rectangles denote the diverging relationship. Because the conflict model differs, the results of depth-first spanning trees are also different for single-lane and multi-lane intersections. 

\begin{figure*}
\begin{center}
    \includegraphics[scale = 0.34]{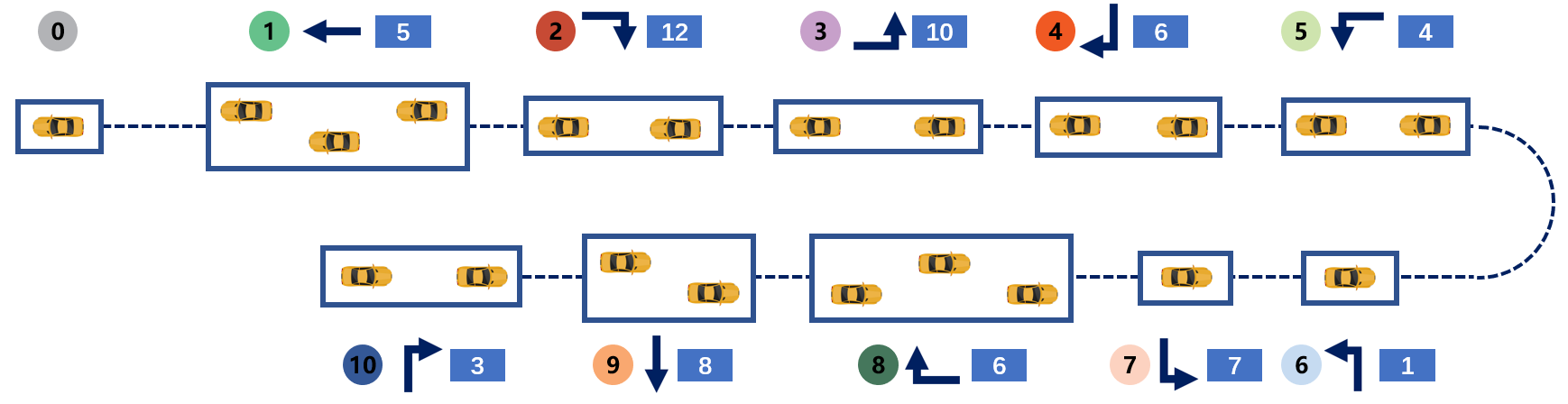}
    \caption{Virtual platoon formed by vehicle groups according to the distance towards the center of the intersection. The numbers in the circles mean the order of these groups based on their distance towards the center of the intersection. The number in the rectangles represent their traffic flow movements, which corresponds to Fig. \ref{tflow}.}
    \label{virtual_platoon}
\end{center}
\end{figure*}

Given an example of virtual platoon where the order of vehicle groups is calculated according to the distance between the first vehicle in the group and the center of the intersection, shown in Fig. \ref{virtual_platoon}. The No. 0 vehicle is a virtual leading vehicle used for generating control input for the vehicle group that is the closest to the center of the intersection. The directed graph is constructed, shown in Fig. \ref{direct2}, to show the conflict relationship of the real and virtual vehicle groups. In Fig. \ref{direct2}, numbers in circles represent vehicle groups given in Fig. \ref{virtual_platoon}. The solid lines show the conflict relationship between vehicle groups based on Fig. \ref{graph2}. Accordingly, the depth-first spanning tree is calculated based on the improved depth-first searching algorithm provided in \cite{Xu2018Distributed}, shown in Fig. \ref{tree2}. The spanning tree shows the order of vehicle groups to pass the intersection. 

The traffic flow movement of the virtual leading vehicle in the virtual platoon is specially defined and in conflict with all the 12 traffic flow movements. In fact, the virtual vehicle conflict with all the other vehicle groups, but Fig. \ref{direct2} omits some of the edges to make itself clearer and easier for understanding.

\begin{figure*}
\begin{center}
    \includegraphics[scale = 0.38]{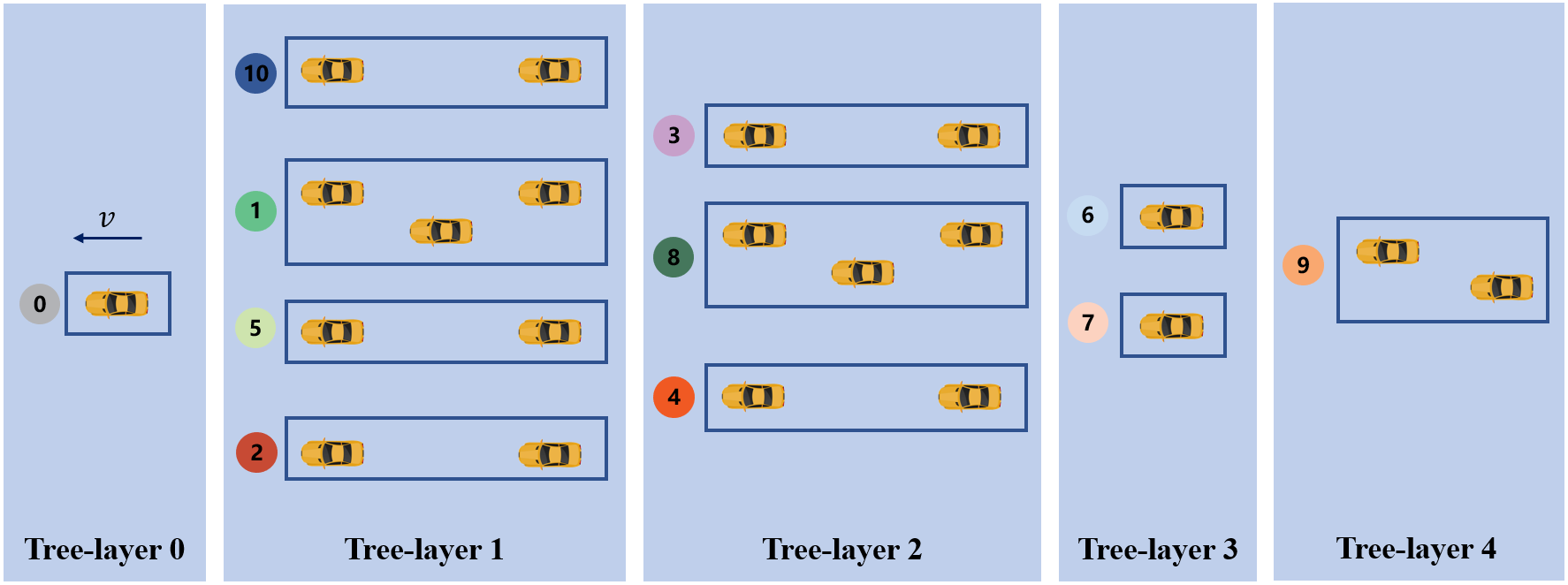}
    \caption{Passing sequence of vehicle groups according to the depth-first spanning tree.}
    \label{treelevel}
\end{center}
\end{figure*}

The vehicle groups given in Fig. \ref{virtual_platoon} and the depth-first spanning tree given in Fig. \ref{tree2} are combined and the passing sequence of the vehicle groups is constructed as shown in Fig. \ref{treelevel}. Vehicle groups that are in the same tree-layer can pass the intersection simultaneously and groups in adjacent tree-layers keep a safe following distance and pass the intersection without collision.

The longitudinal control law of vehicles are given in \eqref{eq_law} and the reference position is given as:
\begin{eqnarray}
\label{eq_xi}
{X_{\text{ref,}i}^k}=X_1^k+x_{gi}\ ,i\not=1
\end{eqnarray}
where ${X_{\text{ref,}i}^k}$ refers to the reference longitudinal position of the $i$-th vehicle in formation $k$, $X_1^k$ refers to the first and also the most forward vehicle in formation $k$, and $x_{gi}$ is defined in \eqref{xy}. 

Equation \eqref{eq_xi} shows that vehicles in a formation follows the most forward vehicle's motion. In order to calculate control input for the most forward vehicle in a formation, the formation set is defined, which is represented as $\mathbb{T}_i$. $\mathbb{T}_i$ is a set that contains the ID of formations that are in tree-layer $i$. Take the spanning tree shown in Fig. \ref{treelevel} as an example, the first formation set $\mathbb{T}_1$ would be $\{1,2,5,10\}$, and $\mathbb{T}_2$ would be $\{3,4,8\}$.

The reference position of the most forward vehicle in a formation is given as:
\begin{eqnarray}
{X_{\text{ref,}1}^k}=X_0+mD_t+\sum_{i=0}^{m-1}(\max_{n\in{\mathbb{T}_i}}(\max{x_{gj}^n}))
\end{eqnarray}
where ${X_{\text{ref,}1}^k}$ refers to the reference position of the most forward vehicle in formation $k$, $X_0$ refers to the position of the virtual leading vehicle, $m$ is the tree-layer that formation $k$ locates, $D_t$ refers to the ideal following gap between two adjacent formations in the virtual platoon, $x_{gj}^n$ refers to the position of the $j$th vehicle in the $n$th formation. When given a certain $n$, $x_{gj}^n$ is equal to $x_{gj}$.

The position of the virtual leading vehicle is given as:
\begin{eqnarray}
X_0={X_{1}^1}-D_t
\end{eqnarray}
which means that the position of the virtual vehicle is defined based on the position of the No.1 vehicle group.
\begin{remark}
In this section, the origin of coordinates is set at the center of the intersection so that the vehicle closer to the center of the intersection has the smaller longitudinal coordinate.
\end{remark}

%
\section{Vehicle regrouping method}
\label{section5}
%

In Section.\ref{section3} and Section.\ref{section4} the multi-lane formation control method for ICVs at one-dimensional straight road and two-dimensional unsignalized intersections are proposed respectively. As is talked in Section.\ref{section4}, vehicles whose target direction is the same are considered as a group. In this section, the vehicle regrouping method is proposed to divide the original formation into subformations in which vehicles have the same turning expectation, e.g., turning left or going straight. The detailed process of the vehicle regrouping method can be seen in Fig. \ref{fig_51}. The process can be divided into the following steps:

\begin{figure*}
\begin{center}
    \includegraphics[scale = 0.35]{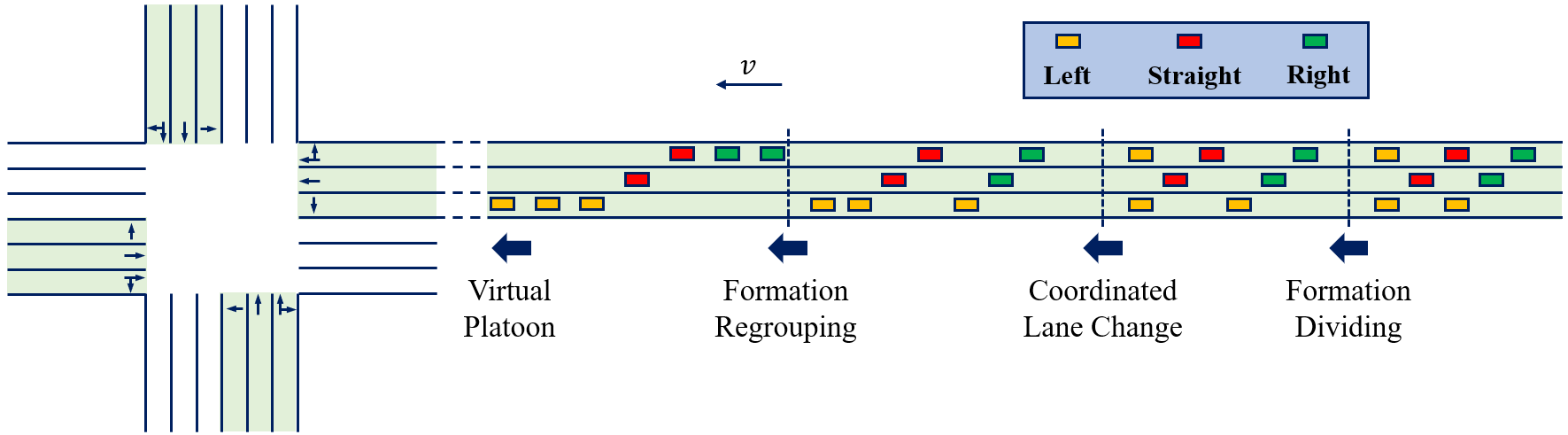}
    \caption{Vehicles' coordinated behavior in the approaching zone and the adjusting zone.}
    \label{fig_51}
\end{center}
\end{figure*}

\begin {itemize}
\item Approaching. Vehicles approach the intersection in multi-lane interlaced formations. The proposed multi-lane formation control method does not consider vehicles' different directions and expected lanes in this approaching zone.
\item Division. When entering the adjusting zone, the formation is divided into smaller subformations according to vehicles' expected lanes. In the dividing process, vehicles don't change lanes and only adjust their reference longitudinal positions in the formation. After the division, vehicles in each subformation have the same directions and different subformations don't intersect longitudinally. In this process, the order of vehicles keeps unchanging.
\item Lane change. Coordinated lane change is performed when subformations have formed. They change to their expected lanes according to the lane assignment. The multi-lane formation switching method proposed in Section.\ref{section3} is adopted here. The upper-level controller calculates target positions and trajectories for vehicles and on-board lower-level controller controls the vehicle to track the trajectory and avoid collision. So, the safety issue is solved by the lower-level controller and the efficiency issue is solve by the upper-level controller. After this process, the subformations would drive on their expected lanes but vehicles inside subformations may still need to change lanes.
\item Formation adjusting. Vehicles adjust their longitudinal and lateral position to form tighter subformations when coordinated lane change has finished. It has been mentioned before that the interlaced geometry is chosen for straight road formation to provided convenience for lane change and formation adjusting. In order to keep the geometric structure of the formation to better connect with the next multi-lane road segment, the interlaced formation structure is adopted.
\item Formation merging. If there is no other conflicting vehicles between two subformations and they are within a desired merging range, they merge to form a new subformation. 
\end {itemize}

The vehicles finish the five steps in the adjusting zone. Then the formations enter the coordinating zone. In this zone, each subformation is considered as a virtual vehicle and will pass the intersection continuously. Then the virtual platoon method can be applied to solve this unsignalized multi-lane intersection problem.

%
\section{Simulation and results}
\label{section6}
%

Simulation experiments in one-dimensional and two-dimensional scenarios are carried out respectively to verify the proposed methods' functions and compare them with reference methods. Simulation environments are built using MATLAB 2018b and SUMO 0.32. Vehicle parameters can be seen in Section.\ref{section2}.


\subsection{One-dimensional formation control}
\label{section6.1}

As is provided in Section.\ref{section3}, the proposed method enables vehicular formation to adjust their geometric shape automatically in the dynamical traffic environment. At first, the proposed method is implemented in the simulation environment to verify its capability, and then the lane assignment method proposed in \cite{21Dao2007Optimized} is chosen to compare the performance in a highway scenario with on-ramps and off-ramps.  The parameters of this part of simulation is shown in Table.\ref{para1}.

The basic idea of the optimal lane assignment method is to assign vehicles to the calculated optimal lanes of the main road at the on-ramps and off-ramps. The goal is to minimize the total traveling time of all the vehicles at each optimizing step. It models the multiple lane assignment problem into a linear programming problem. By taking the percentage of vehicles to be assigned to each lane as independent variable, while maintaining the number of vehicles in a certain lane less than the biggest lane capacity, the authors solve the linear programming problem and get the best result of lane assignment.

\begin{table}[htbp]
\centering
\caption{Simulation parameters part 1}
\label{para1}
\begin{tabular}{lll}
\toprule
Simulation time step                                   &   $\Delta t$              & 0.04 s \\ 
Length of the highway segment                          &   $L_{segment}$   & 2000 m \\ 
Length of the on-ramp merging segment                  &   $L_{on}$        & 100 m \\ 
Length of the off-ramp diverging segment               &   $L_{off}$       & 50 m \\ 
Lane witdh                                                            &   $W_{lane}$       & 4 m \\ 
Weight of longitudinal distance           &   $w_{1}$       & 1 \\
Weight of lateral lane change number     &   $w_{2}$       & 10 \\
\\
Desired speed on the ramp                              &   $v_{r,des}$              & 15 m/s \\ 
Maximum speed on the ramp                              &   $v_{r,max}$              & 25 m/s \\ 
Minimum speed on the ramp                              &   $v_{r,min}$              & 5 m/s \\ 
\\
Desired speed on the left three lanes                  &   $v_{m,des}$              & 25 m/s \\ 
Desired speed on the fourth lane     &   $v_{m,des}'$              & 20 m/s \\ 
Maximum speed on the main road                         &   $v_{m,max}$              & 30 m/s \\ 
Minimum speed on the main road                         &   $v_{m,min}$              & 15 m/s \\ 
\\
Maximum acceleration                                   &   $a_{max}$              & 3 m/$\text{s}^2$ \\ 
Minimum acceleration                                   &   $a_{min}$              & -6 m/$\text{s}^2$ \\ 
\\
Desired formation gap on the main road                 &   $L_s$              & 20 m \\ 
\bottomrule  
\end{tabular}
\end{table}

\begin{figure}
\vspace{1mm}
\begin{center}
    \subfigure[]{
    \includegraphics[scale = 0.225]{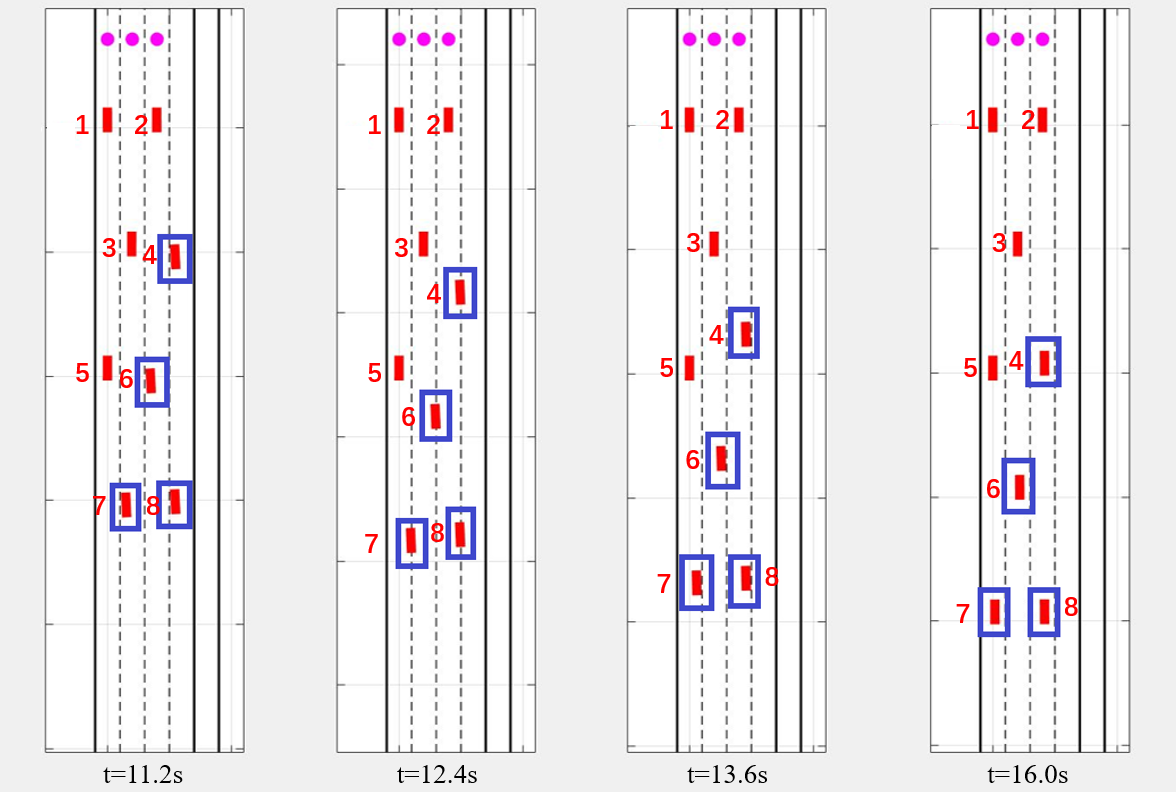}
    \label{fig_simu_lanenum}}
\hspace{1mm}
    \subfigure[]{
    \includegraphics[scale = 0.225]{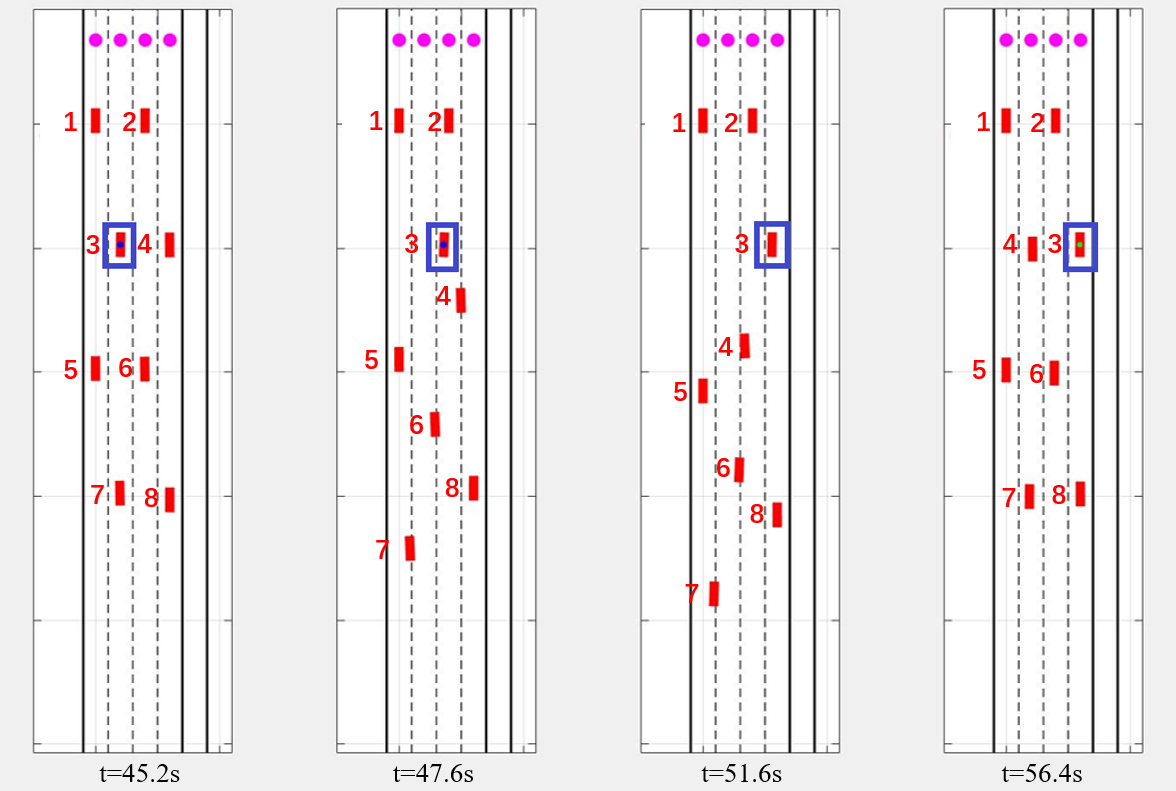}
    \label{fig_simu_lanechange}}

    \subfigure[]{
    \includegraphics[scale = 0.225]{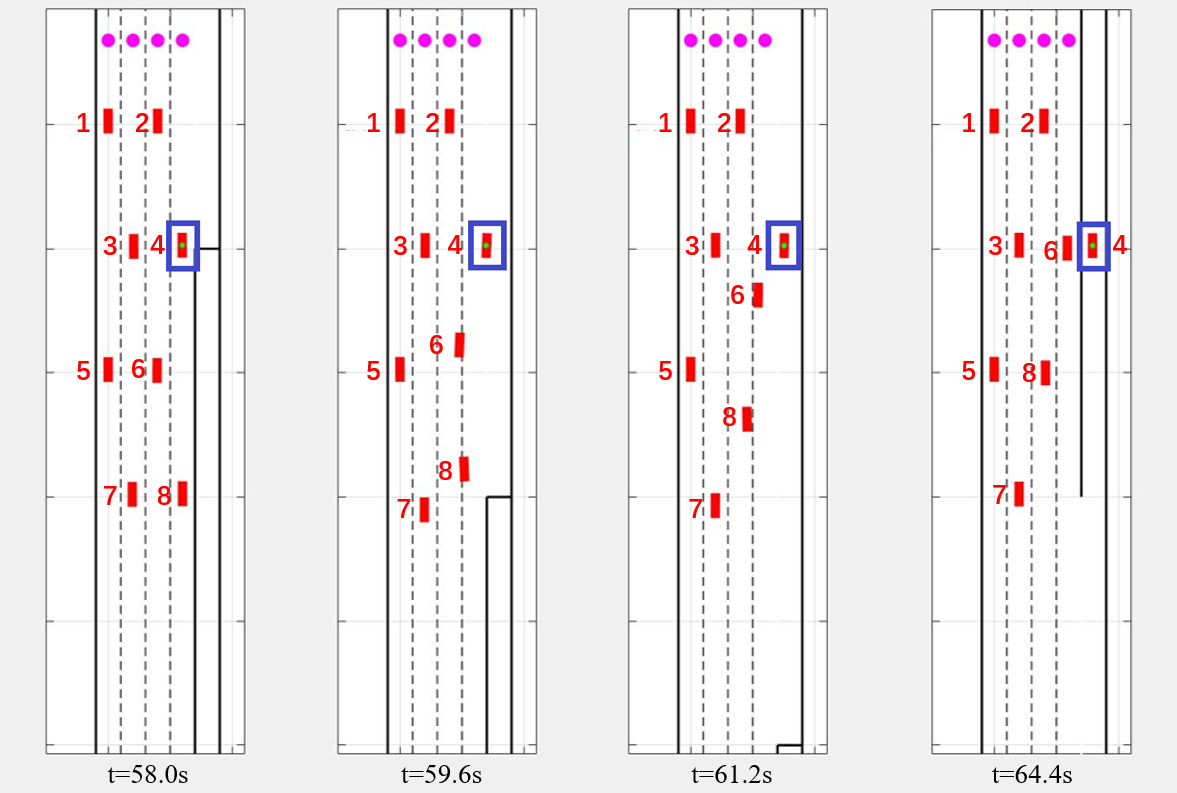}
    \label{fig_simu_offramp}}
\hspace{1mm}
    \subfigure[]{
    \includegraphics[scale = 0.225]{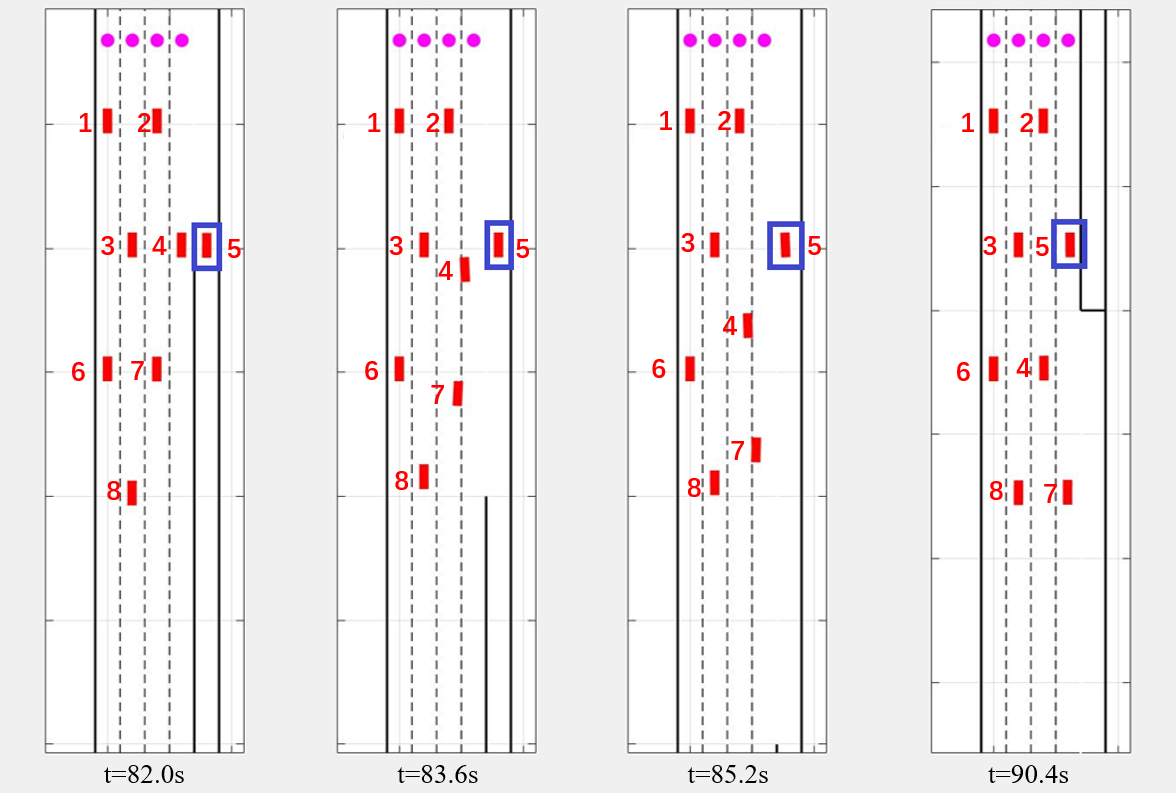}
    \label{fig_simu_onramp}}
    \caption{Simulation results of the proposed method. (a) shows the process of formation forming when the number of lanes changes from four to three; (b) shows the process of inner-formation lane changing; (c) shows the process of vehicle leaving formation at an off-ramp and (d) shows the process of new vehicle joining formation at an on-ramp.}
    \label{fig_simu_own}
\end{center}
\end{figure}

First of all, simulation under simple traffic environment is carried out. The results of implementation of the proposed method are shown in Fig. \ref{fig_simu_own}, where four shots of each scenario were selected. In Fig. \ref{fig_simu_own}, the red rectangles represent vehicles and the black solid and dashed lines represent lanes. There are five lanes in total. The left four lanes are common roads and the right most one is a ramp. The magenta circle on the top means that the lane is drivable.

In order to compare the proposed method with the reference method, a highway simulation environment with four regular lanes and one lane on the right to represent on-ramps and off-ramps is built. The total length of the highway is 10000 m with five on-ramps and five off-ramps. On-ramps are set every 2000 m and the first on-ramp is at the start of the highway. Off-ramps are set also every 2000 m and the first off-ramps is set at 1600 m after the first on-ramp. The repeatable basic segment of the simulation environment is shown in Fig. \ref{fig_envir}. In order to maintain the smoothness of traffic flow, the start of the 10000-meter road is connected with its end. All vehicles are generated at the five on-ramps randomly. Once a new vehicle is generated, it chooses its exit among the five off-ramps randomly. It should be noted that when using the proposed method in the simulation, the fourth lane, which is a transition lane, is open for vehicles only those who are going to leave or join a formation, so the formation is built on the left three lanes.

\begin{remark}
In the simulation, the fourth lane is set as a transition lane at the on-ramps and the off-ramps. This is because the vehicles that have just merged into the main road from on-ramps or are going to leave the main road at the next off-ramp usually have lower speed than others. In order to avoid collision and keep high traffic efficiency, the vehicular formations are built only on the left lanes at the ramp areas.
\end{remark}

\begin{figure*}
\vspace{5mm}
\begin{center}
    \subfigure[Simulation results of the proposed method]{
    \includegraphics[scale = 0.55]{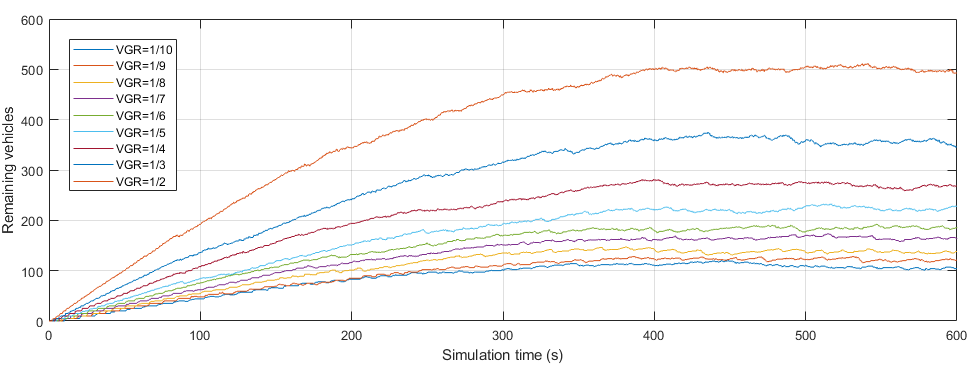}
    \label{fig_remain_own}}
    \subfigure[Simulation results of the reference method]{
    \includegraphics[scale = 0.55]{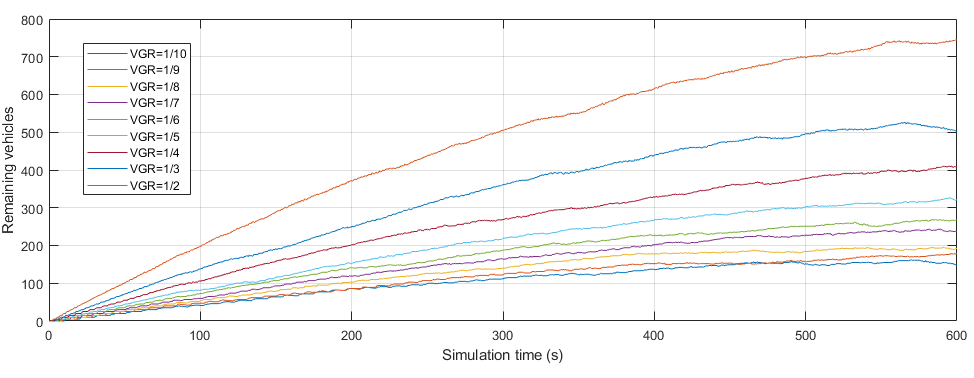}
    \label{fig_remain_ref}}
    \caption{Results of the number of vehicles remaining on the roads during $600s$ at different VGR.}
    \label{fig_remain}
\end{center}
\end{figure*}

The vehicle generating rate (VGR) is changed at on-ramps to watch the changing trend of the number of vehicles remaining on the road.
It can be found that using the proposed method, which can fully use the capacity of the lanes, can significantly reduce the number of remaining vehicles, which is about 30\% less at the same VGR compared with the reference method. It indicates that the proposed method can reduce the load of highways. The result of the proposed method is shown in Fig. \ref{fig_remain_own} and the result of the reference method is shown in Fig. \ref{fig_remain_ref}.

\begin{figure*}
\begin{center}
    \subfigure[Vehicle distribution of the proposed method]{
    \includegraphics[scale = 0.32]{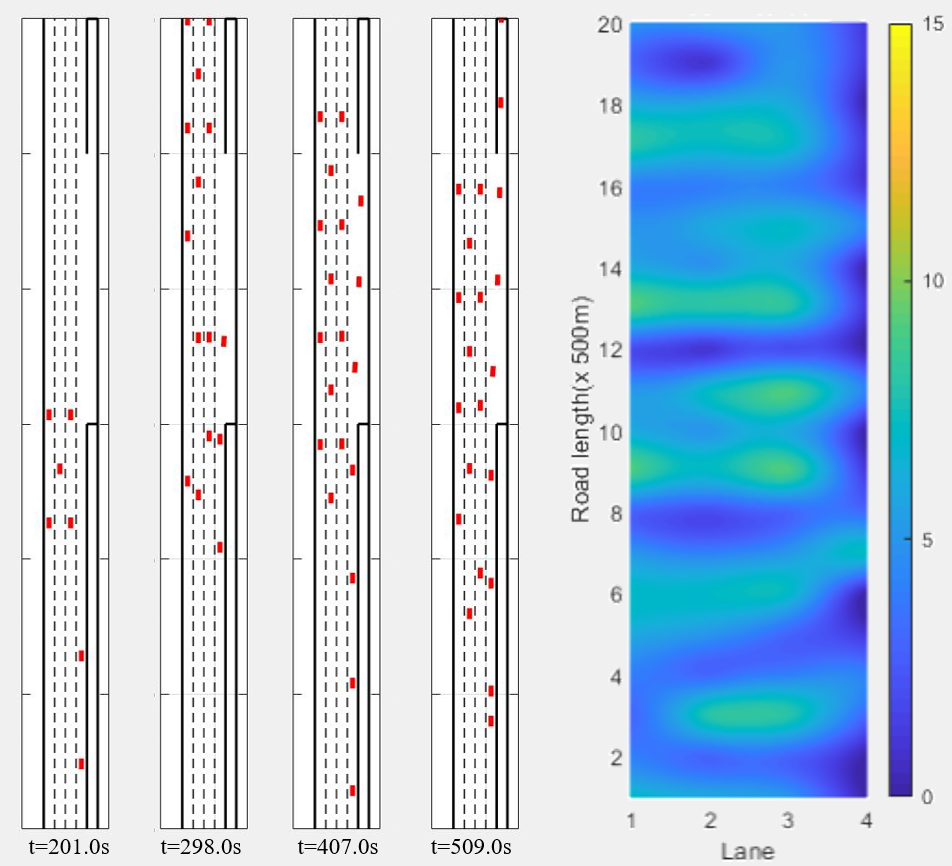}
    \label{fig_heat_1}}
\hspace{2mm}
    \subfigure[Vehicle distribution of the reference method]{
    \includegraphics[scale = 0.32]{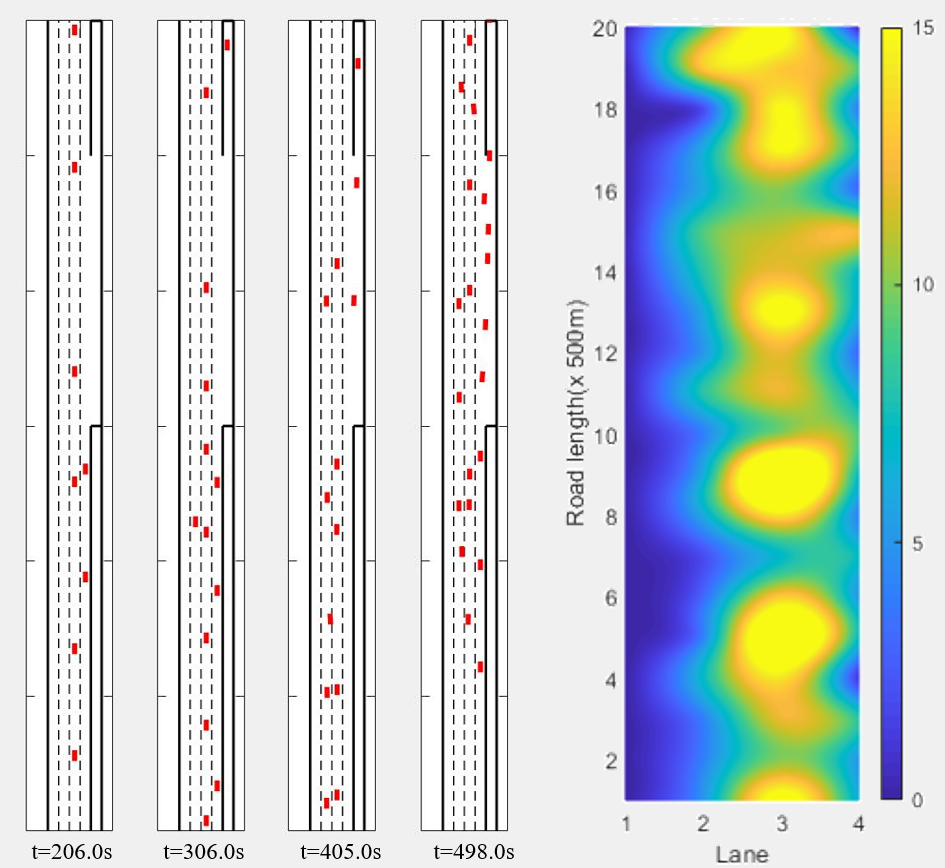}
    \label{fig_heat_2}}
    \caption{Snapshots of the simulation at an off-ramp and heatmap of the two methods.}
    \label{fig_shot}
\end{center}
\end{figure*}

In order to show the distribution of vehicles during the simulation, the window is fixed at a road segment around an off-ramp. Fig. \ref{fig_heat_1} shows the distribution of vehicles around an off-ramp at the simulation time around 200 s, 300 s, 400 s and 500 s respectively when using the proposed method. Correspondingly, Fig. \ref{fig_heat_2} shows the distribution of vehicles of the reference method. Besides, the heat maps are used to show the distribution of vehicles at simulation time 600 s and at the VGR $1/3$ vehicle/s inFig. \ref{fig_heat_1} and Fig. \ref{fig_heat_2}. They are constructed to show the density of vehicles. The number of vehicles remaining on the main road is counted every 500 m. The brighter the color is, the more crowded the vehicle is. The darker the color is, the sparser the vehicle is. It can be seen from the heatmap that the main difference between the two methods is that the proposed method makes better use of capacity of all the lanes while the reference method may cause some congestion in some certain lanes.

\begin{figure}
\begin{center}
    \subfigure[]{
    \includegraphics[scale = 0.21]{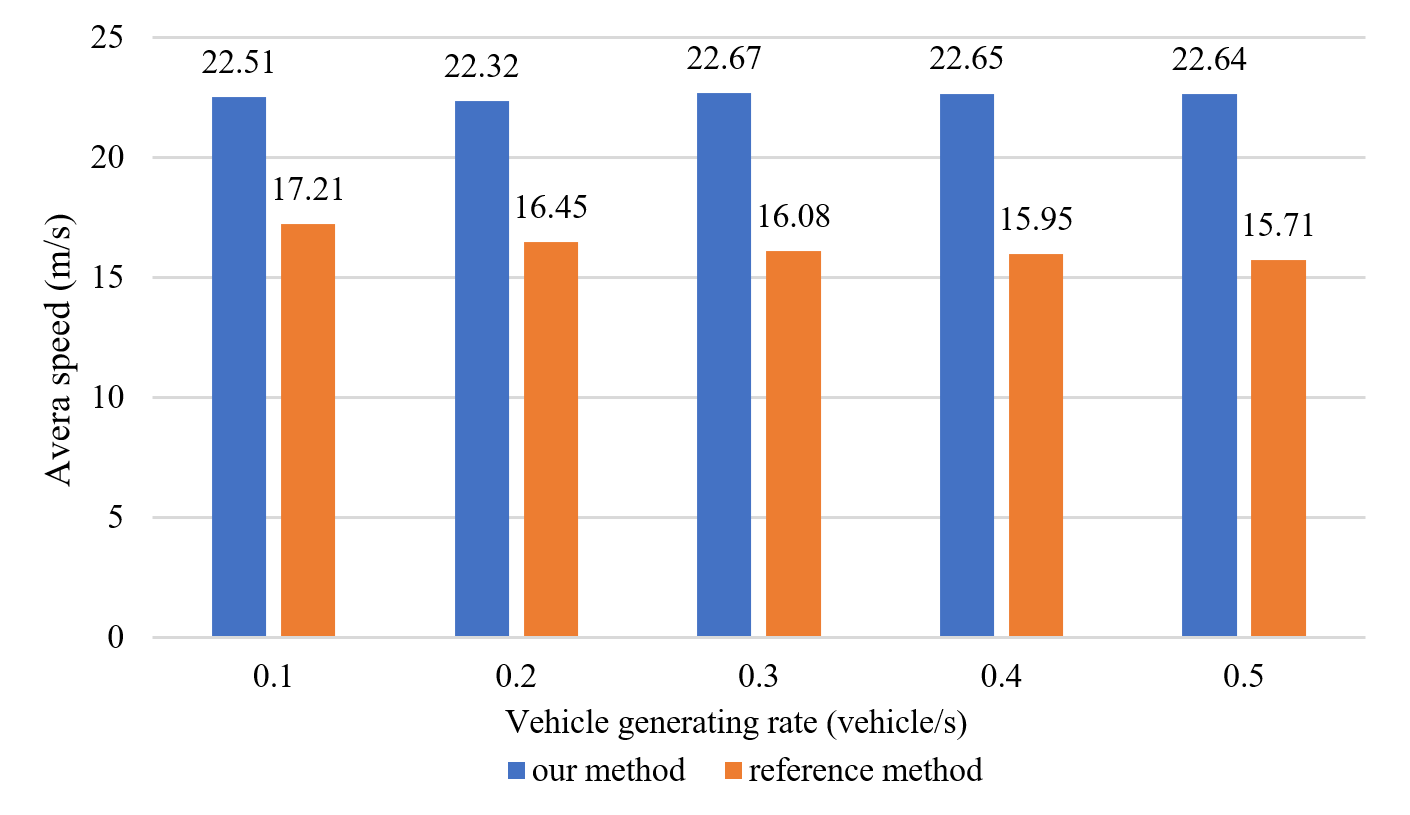}
    \label{fig_result_speed}}
    \subfigure[]{
    \includegraphics[scale = 0.21]{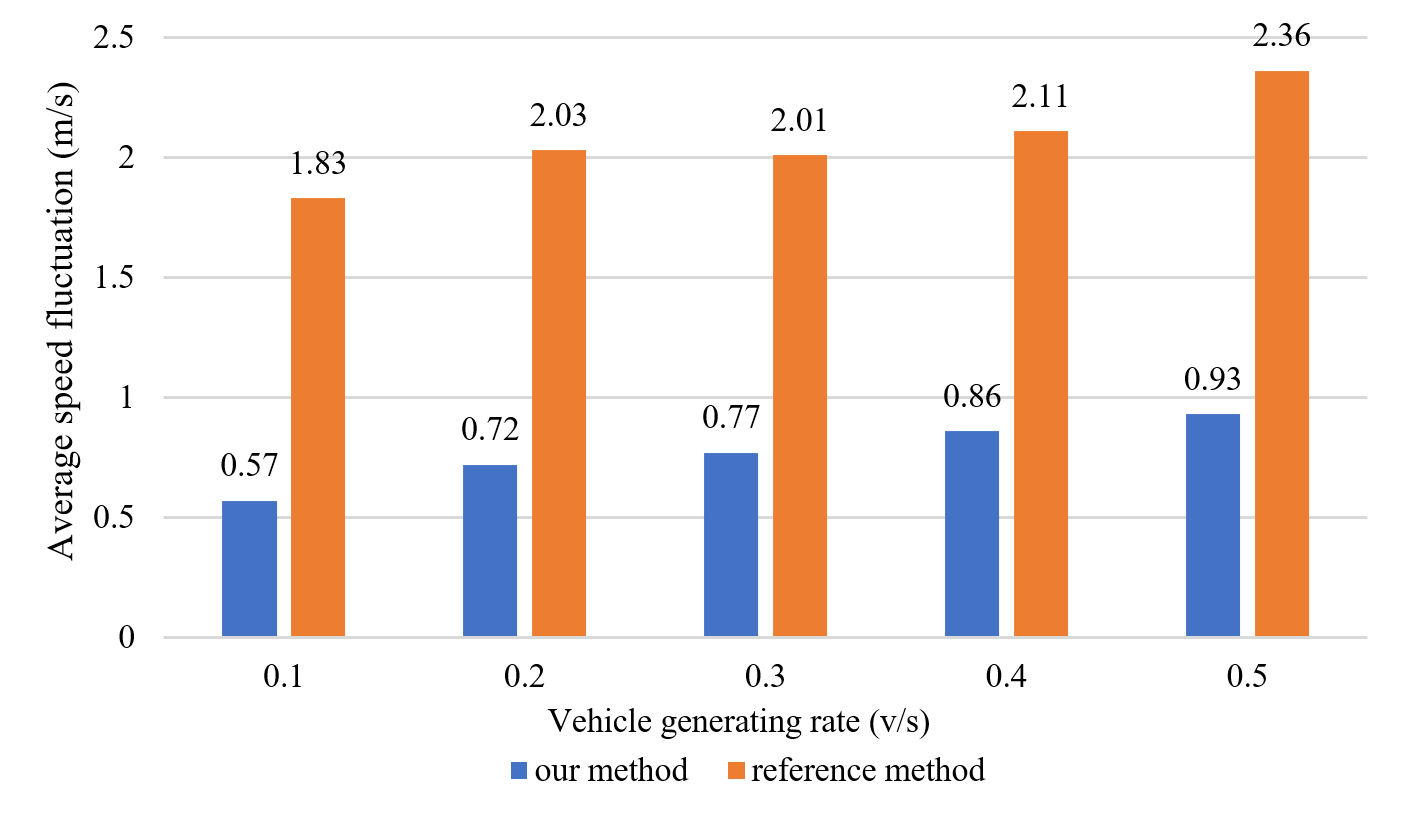}
    \label{fig_result_acc}}
    \caption{Average speed and acceleration of all the vehicles under different VGR. The red line shows the result of our method and the blue line shows the result of the reference method.}
    \label{fig_result2}
\end{center}
\end{figure}

The average speed and speed fluctuation of all the vehicles are calculated and the results are shown in Fig. \ref{fig_result2}. It can be found in Fig. \ref{fig_result_speed} and Fig. \ref{fig_result_acc} that using the proposed method can achieve a 30\% higher average speed and a 60\% lower average speed fluctuation compared with the result of the reference method. It indicates that the proposed method can do better in traffic efficiency and significantly reduce frequent accelerating and decelerating to achieve better fuel economy.

\subsection{Two-dimensional unsignalized intersection control}

The simulation environment for the two-dimensional unsignalized intersection control is built with the same lane assignment and lane numbers as is shown in Fig. \ref{figure_single_inter}. The Vehicle regrouping method is applied in the adjusting zone to form subgroups and perform coordinated lane change to guide vehicles to their expected lanes. The multi-lane virtual platoon method is applied in the cooperating zone taking the subgroups formed by the Vehicle regrouping method as nodes in the spanning tree to calculate conflict-free passing sequence. The parameters of this part of simulation is shown in Table.\ref{para2}.

\begin{table}[htbp]
\centering
\caption{Simulation parameters part 2}
\label{para2}
\begin{tabular}{lll}
\toprule
Simulation time step                                             &   $\Delta t$              & 0.04 s \\
Radius of the coordinating zone                                  &   $R_1$           & 300 m \\ 
Radius of the adjusting zone                                     &   $R_2$           & 500 m \\ 
Radius of the approaching zone                                   &   $R_3$              & 800 m \\ 
\\          
Desired speed of the virtual leading vehicle                     &   $v_{0,des}$              & 10 m/s \\ 
Maximum speed                                                    &   $v_{max}$              & 20 m/s \\ 
Minimum speed                                                    &   $v_{min}$              & 0 m/s \\ 
Maximum acceleration                                             &   $a_{max}$              & 3 m/$\text{s}^2$ \\ 
Minimum acceleration                                             &   $a_{min}$              & -6 m/$\text{s}^2$ \\ 
\\
Desired gap outside the coordinating zone      &   $L_s$              & 15 m \\ 
Desired gap in the coordinating zone          &   $L_s$              & 9 m \\ 
\bottomrule  
\end{tabular}
\end{table}

\begin{figure*}
\begin{center}
    \includegraphics[scale = 0.42]{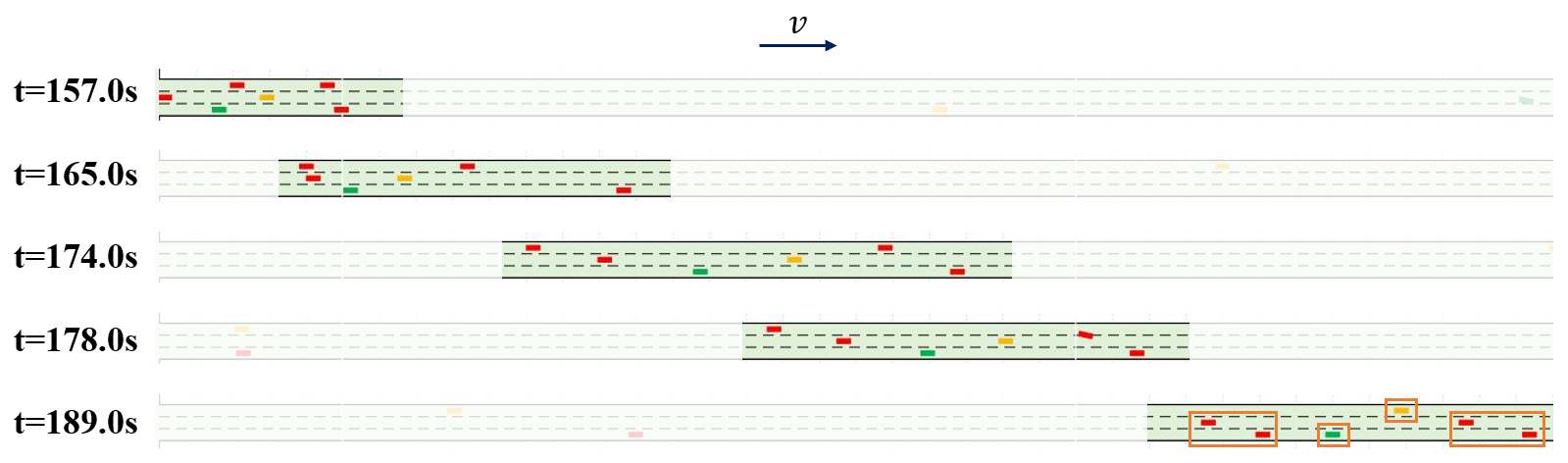}
    \caption{Simulation process of the Vehicle regrouping method at different simulation time. The subformations are marked with re hollow rectangles.}
    \label{fig_di}
\end{center}
\end{figure*}

Vehicles are generated at the entrance of four legs. The generating rate of vehicles are set to control the throughput of the intersection. Vehicles form multi-lane interlaced formation in the approaching zone. When the first vehicle of a formation enters the adjusting zone, the formation then performs regrouping process. Vehicles finish lane change in the adjusting zone and form tighter subformations in their expected lanes. The process of regrouping is shown in Fig. \ref{fig_di}. Because no lane changes need to be done after vehicles entering the coordinating zone, vehicles drive with smaller following distance. When subgroups enter the coordinating zone, they are taken as nodes of the virtual platoon. Conflict-free passing sequence is then calculated using the depth-first spanning searching algorithm.

\begin{figure*}
\begin{center}
    \subfigure[Multi-vehicle virtual platoon method]{
    \includegraphics[scale = 0.24]{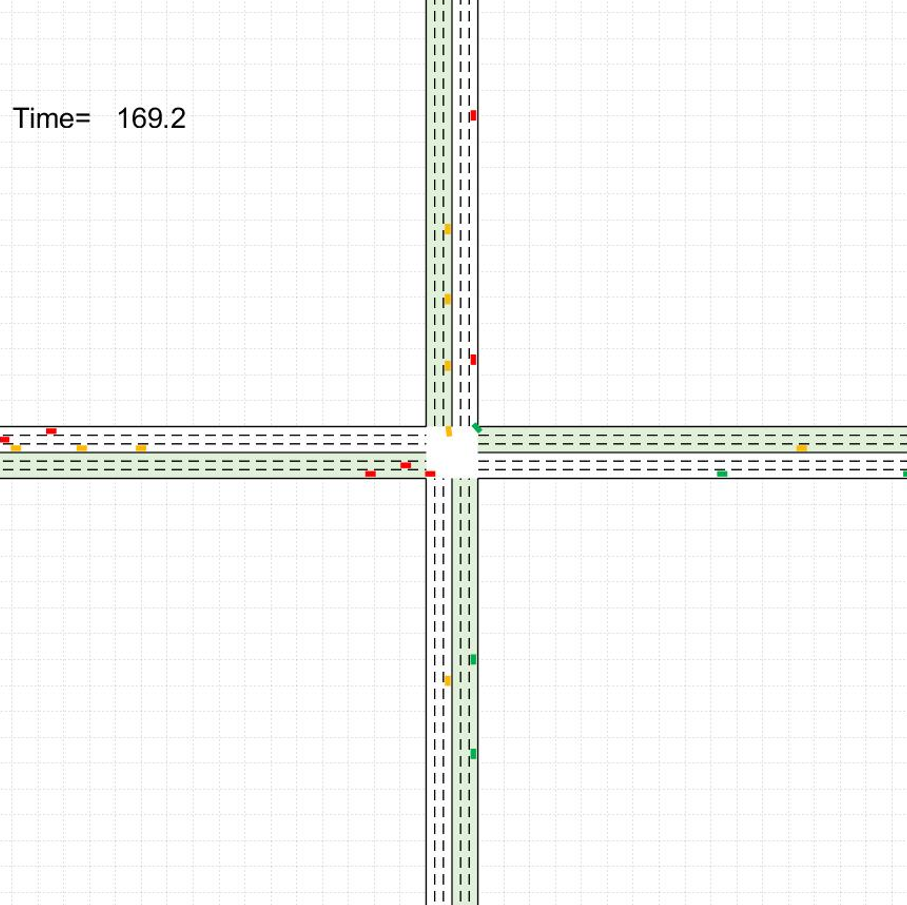}
    \label{fig_shot_multi}}
    \subfigure[Single-vehicle virtual platoon method]{
    \includegraphics[scale = 0.24]{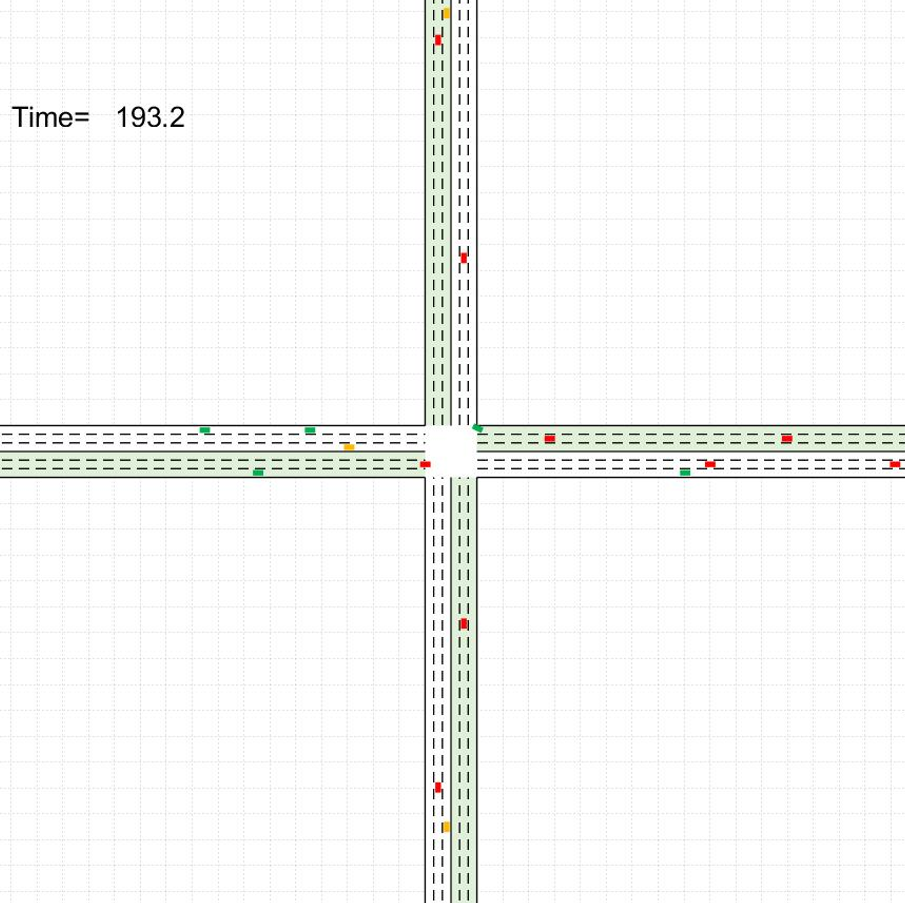}
    \label{fig_shot_single}}
    \subfigure[Reservation-based method]{
    \includegraphics[scale = 0.24]{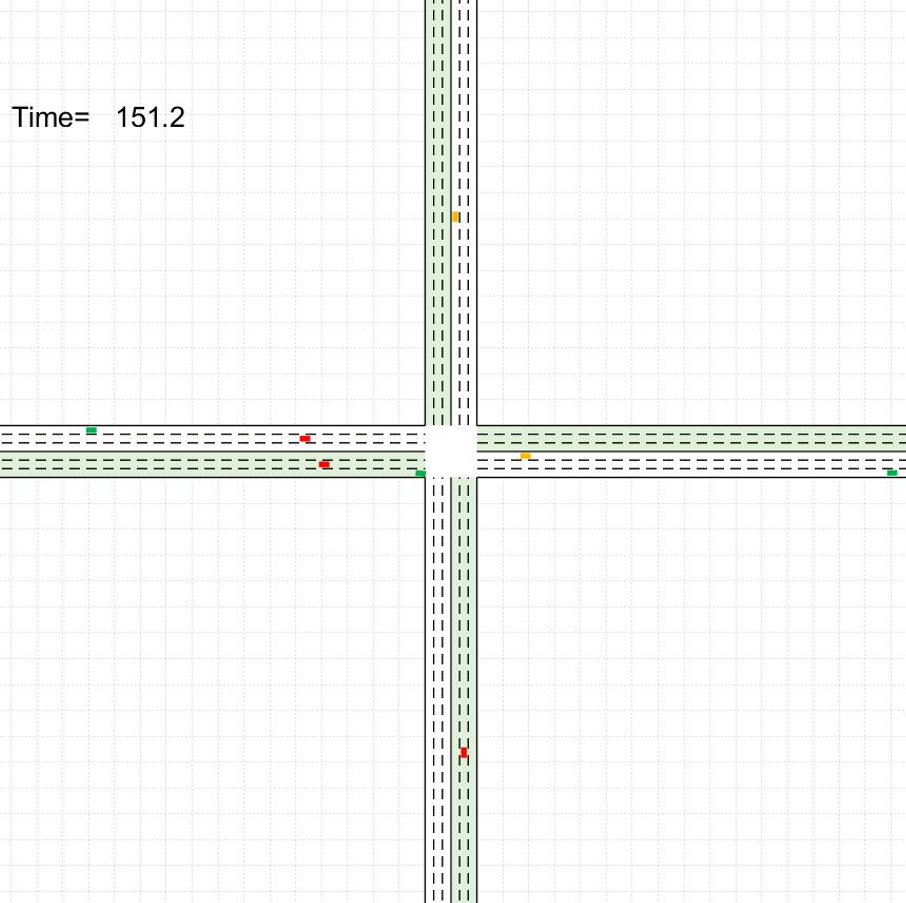}
    \label{fig_shot_res}}
    \caption{Snapshots of the intersection of different methods.}
    \label{fig_in}
\end{center}
\end{figure*}

The reservation-based method and the classic virtual platoon method are chosen as reference methods. In the reservation method, vehicles pass the intersection one by one according to the ``First Come First Serve" principle. In the classic virtual platoon method, each single vehicle is taken as a node in the spanning tree. The snapshots of the simulation using the above three methods are shown in Fig. \ref{fig_in}.

\begin{figure*}
\begin{center}
    \subfigure[]{
    \includegraphics[scale = 0.31]{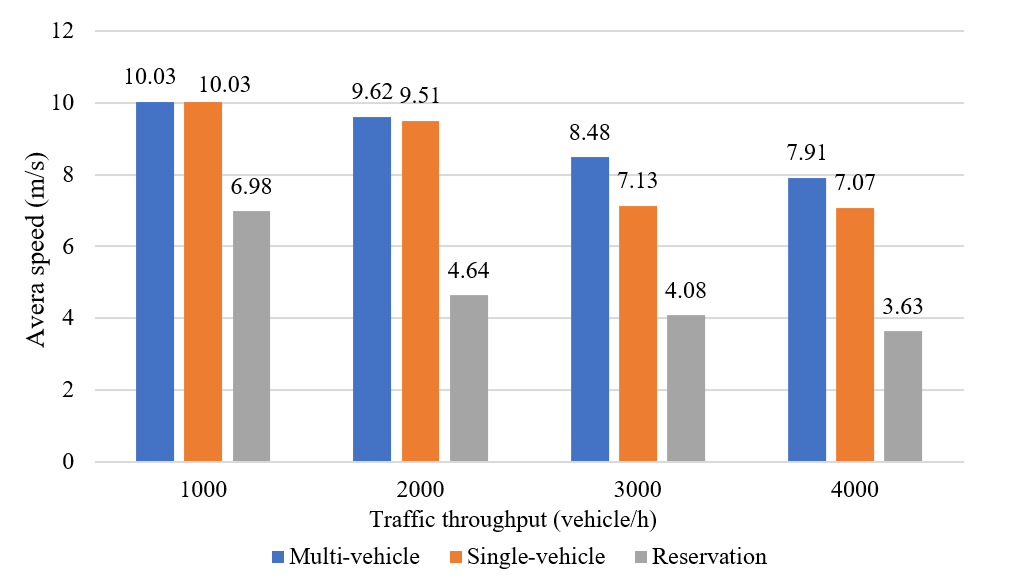}
    \label{fig_as2}}
    \subfigure[]{
    \includegraphics[scale = 0.31]{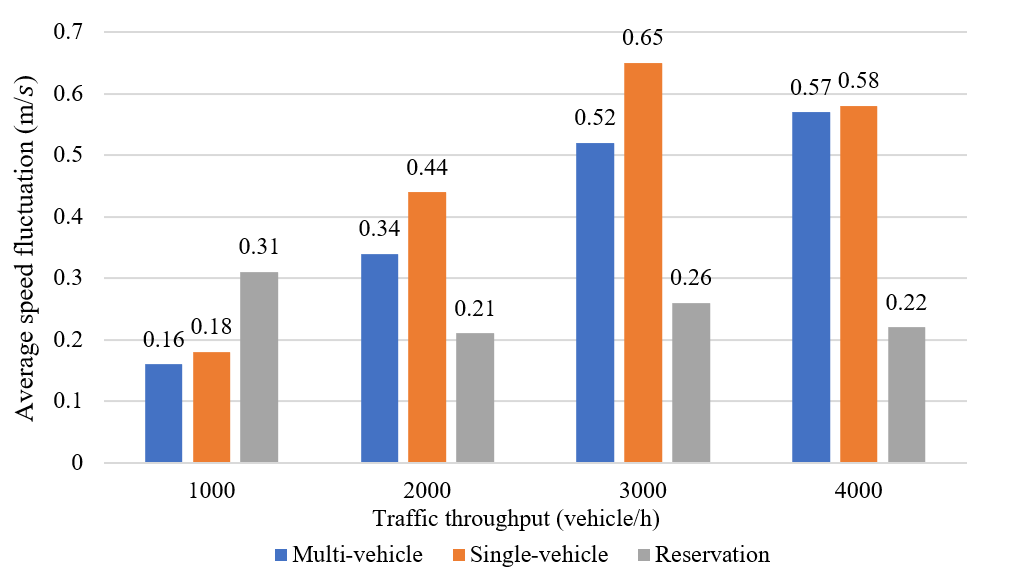}
    \label{fig_aa2}}
    \caption{Average speed and acceleration of all the vehicles of the multi-lane unsignalized intersection under different traffic throughput.}
    \label{fig_re2}
\end{center}
\end{figure*}

Fig. \ref{fig_in} indicates that compared with the reservation-based method, the virtual platoon method (either multi-vehicle or single-vehicle) can achieve a higher vehicle density near the intersection because vehicles that are not in conflict can pass the intersection simultaneously. Besides, compared with the single-vehicle platoon method, the proposed multi-vehicle platoon method can further increase vehicle density because vehicles with the same target direction can drive with a closer distance in a formation. Average speed and average speed fluctuation are chosen to evaluate traffic efficiency and fuel economy respectively. The vehicle generating rate is changed for different traffic throughout to get different performance. The results are shown in Fig. \ref{fig_re2}. 

It can be seen from Fig. \ref{fig_as2} that at low traffic throughput (lower than 2000 vehicle/h), both the multi-vehicle and single-vehicle virtual platoon method maintain an average speed which is close to the expected speed (10 m/s). When the traffic throughput becomes higher, the average speed goes down and the multi-vehicle method always exceeds the single-vehicle method. The average speed of the reservation-based method also goes down when the traffic throughput becomes higher, and is always lower than the two virtual platoon methods. This indicates that the virtual platoon method can improve traffic efficiency compared with the reservation-based method and the multi-vehicle method can achieve a better result than the single-vehicle method. 

The result of average speed fluctuation is shown in Fig. \ref{fig_aa2}. It can be seen that the reservation-based method always keeps a low fluctuation, because when using this method, vehicles just have to decelerate or brake until they reserve successfully for their passing time so that they don't need to frequently accelerate or decelerate. The speed fluctuation of the multi-vehicle virtual platoon method is always lower than the single-vehicle method which indicates that the multi-vehicle method can achieve better fuel economy than the single-vehicle method. Although the reservation-based method keeps the lower speed fluctuation at higher traffic throughput, it is clear that compared with the loss in fuel economy, the improvement in traffic efficiency is much more remarkable. It's interesting to find that the speed fluctuation of the single-vehicle method goes down at the highest traffic throughput. This is because that at such a high traffic throughput, many vehicles brake and stop for a long time. The speed of these stopping vehicles don't change so that the statistical average speed fluctuation becomes lower.

%
\section{Conclusion}
\label{section7}
%

In this paper, a unified multi-vehicle formation control framework is proposed for Intelligent and Connected Vehicles (ICVs). In the framework, groups of vehicles drive as formations with flexible geomerty to improve driving safety and traffic efficiency. Vehicles may join, leave or perform coordinated behavior with other vehicles in a formation. The framework is scenario-insensitive and can be a general solution for the vehicle control problem in the intelligent transportation system.

Firstly, the vehicular formation is built on multi-lane roads. Different formation geometries are analysised and the interlaced structure is mathematically modelized to improve driving safety while making full use of the lane capacity. The cost function is specially designed and the assignment problem for vehicles and target positions is modelized and solved using Hungarian Algorithm to improve the flexibility of the method. B$\acute{\text{e}}$zier curves are adopted to guide vehicles to their assigned target positions.

Secondly, the scenario is extended to multi-lane unsignalized intersections. Conflict relationship between vehicular formations is studied and modelized.  Conflict-free passing sequence is calculated using the depth-first spanning algorithm. Vehicle controller is designed for vehicles based on their relative positions in the spanning tree.

Then, the vehicle regrouping method is proposed to connect the one-dimensioanl and two-dimensional scenarios. Formations driving on the multi-lane raods are divided into subformations according to their target directions and expected lanes. The subformations are then taken as the input for the conflict-free passing method in the intersections.

Finally, simulations are carried out to validate the performance of the proposed framework. The results prove that the proposed multi-vehicle formation control framework can cover both one-dimensional and two-dimensional scenarios. Besides, compared with existing methods, the proposed framework has better performance in traffic efficiency and fuel economy while maintaining vehicles driving safely and in order.

\section{Acknowledgments}\label{sec11}

This work was supported by National Key Research and Development Program of China (2019YFB160080), National Natural Science Foundation of China (51605245), and Intel Collaborative Research Institute Intelligent and Automated Connected Vehicles.

\bibliographystyle{iet}
\bibliography{thesis}

\end{document}